\documentclass[10pt,preprint]{aastex}

\begin{document}

\title{Lightcurves of Type Ia Supernovae from Near the Time of Explosion}

\shorttitle{Near Explosion LCs of SNe Ia}
\shortauthors{Garg et al.}

\author{Arti Garg\altaffilmark{1}}
\email{artigarg@fas.harvard.edu}
\author{Christopher W. Stubbs\altaffilmark{1}, Peter Challis\altaffilmark{1}, W.~Michael~Wood-Vasey\altaffilmark{1}, and St{\'e}phane Blondin\altaffilmark{1}}
\affil{Department of Physics and Harvard-Smithsonian Center for Astrophysics \\
Harvard University, Cambridge MA USA}
\author{Mark E. Huber\altaffilmark{1}, Kem Cook\altaffilmark{1}, and Sergei Nikolaev}
\affil{Lawrence Livermore National Laboratory}
\author{Armin Rest\altaffilmark{2}, R. Chris Smith, Knut Olsen, Nicholas~B.~Suntzeff, and Claudio Aguilera }
\affil{Cerro Tololo Inter-American Observatory/NOAO\altaffilmark{3}}
\author{Jose Luis Prieto\altaffilmark{1}}
\affil{Ohio State University}
\author{Andrew Becker\altaffilmark{1} and Antonino Miceli}
\affil{University of Washington}
\author{Gajus Miknaitis\altaffilmark{1}}
\affil{Fermi National Accelerator Laboratory}
\author{Alejandro Clocchiatti\altaffilmark{1}, Dante Minniti, and Lorenzo Morelli\altaffilmark{1}}
\affil{P. Universidad Catolica}
\author{Douglas L. Welch\altaffilmark{1}}
\affil{McMaster University}

\altaffiltext{1}{Visiting Astronomer, Cerro Tololo Inter-American
Observatory, National Optical Astronomy Observatory, which is operated
by the Association of Universities for Research in Astronomy,
Inc. (AURA) under cooperative agreement with the National Science
Foundation}
\altaffiltext{2}{Goldberg fellow}
\altaffiltext{3}{Cerro Tololo Inter-American Observatory, National
Optical Astronomy Observatory, which is operated by the Association of
Universities for Research in Astronomy, Inc. (AURA) under cooperative
agreement with the National Science Foundation}

\begin{abstract}
We present a set of 11 type Ia supernova (SN~Ia) lightcurves with
dense, pre-maximum sampling. These supernovae (SNe), in galaxies
behind the Large Magellanic Cloud (LMC), were discovered by the
SuperMACHO survey.  The SNe span a redshift range of z = 0.11 --
0.35. Our lightcurves contain some of the earliest pre-maximum
observations of SNe~Ia to date.  We also give a functional model that
describes the SN~Ia lightcurve shape (in our $VR$-band).  Our function
uses the ``expanding fireball'' model of Goldhaber et al. (1998) to
describe the rising lightcurve immediately after explosion but
constrains it to smoothly join the remainder of the lightcurve.  We
fit this model to a composite observed $VR$-band lightcurve of three
SNe between redshifts of 0.135 to 0.165.  These SNe have not been
K-corrected or adjusted to account for reddening.  In this redshift
range, the observed $VR$-band most closely matches the rest frame
$V$-band.  Using the best fit to our functional description of the
lightcurve, we find the time between explosion and observed $VR$-band
maximum to be 17.6$\pm$1.3({\it stat})$\pm$0.07({\it sys})~ rest-frame days
for a SN~Ia with a $VR$-band $\Delta m_{-10}$ of 0.52{\it mag}.  For the
redshifts sampled, the observed $VR$-band time-of-maximum brightness
should be the same as the rest-frame $V$-band maximum to within
1.1~rest-frame days.
\end{abstract}

\keywords{surveys---supernovae: general---Magellanic Clouds---{\it
Facilities:} \facility{Blanco ()}, \facility{Magellan:Baade ()},
\facility{Magellan:Clay ()}}

\section{Introduction}
\subsection{Rise-time as a tool to discriminate between SN~Ia explosion models}
The realization that type Ia supernovae (SNe~Ia) can be used as
standardizable candles (Phillips 1993; Riess et al. 1995, 1996; Hamuy
et al. 1996, 1996b) led to an explosion in SN~Ia science.  Surveys to
test the Hubble Expansion Law at larger distances found that rather than
exhibiting a constant or decelerating expansion rate, the Universe has
an accelerating expansion (Riess et al.  1998, Perlmutter et
al. 1999).  The consensus explanation for the accelerating expansion
is a negative pressure, or dark energy, permeating the Universe.
Today many teams are working to use SNe~Ia as standard candles to
better constrain the properties of dark energy (ESSENCE, Matheson et
al. 2005; SCP, Kowalski et al. 2005; SNLS, Astier et al. 2006).  While
the methods to standardize the SN~Ia luminosity vary, the
interpretation of all their results rely to varying degrees on the
basic assumption that SNe~Ia belong to a single-parameter family.

Methods of standardizing SN~Ia luminosity distance using the
post-maximum lightcurve shape have proven successful when verified
against other standard candles such as Cepheids (Suntzeff et al. 1999,
Gibson et al. 2000).  These results do not necessarily indicate the
existence of a single-parameter family of progenitors, only that the
behavior of SNe~Ia post-maximum is similar.  Still, the most widely
considered SN~Ia progenitors are carbon-oxygen (C-O) white dwarfs in
binary systems.  Even accepting these systems as the progenitors,
questions remain concerning the mechanism and progression of the
explosion.  Many competing theories (see Hillebrandt \& Niemeyer 2000
and references therein) predict roughly the same post-maximum behavior
and vary only in the prediction of the pre-maximum, or rising,
lightcurves and spectra.  Understanding the explosion mechanism may
help us better understand how the population of SNe~Ia, and their
progenitors, evolves over cosmological time.  Many explosion models
are sensitive to progenitor element abundances which may vary
depending on the environment.  Combining existing information about
the differences between low- and high-z stellar populations and
galaxies with a more accurate model of the SN~Ia explosion mechanism
will help more tightly constrain the impact of evolution on SN~Ia
lightcurve shape.  Discriminating between competing explosion models,
however, requires lightcurve coverage close to the time of explosion
which has been scarcely available.

The reasons for the lack of early pre-maximum lightcurve coverage are
many-fold.  Some SNe~Ia searches rely on a search-and-follow method
where SNe are discovered and then followed by another, larger
telescope.  Discovery often occurs near maximum brightness, and dense
pre-maximum temporal coverage is not available.  Other surveys,
similar to SuperMACHO, revisit the same fields every few days,
obtaining consistent temporal coverage over the entire lightcurve.
These data sets have better pre-maximum coverage but still do not
generally provide densely-sampled pre-maximum lightcurves.  In order
to maximize the number of fields observed, most surveys use a long,
multi-day gap between observations which is sufficient to standardize
the post-maximum behavior but often misses the earliest portion of the
rise.  For higher-z SNe where the multi-day gap between observations
translates to a shorter gap in the SN's rest-frame, the earliest
portion of the rise is often too faint to be observed.  As described
more completely below, the SuperMACHO data avoid these two pitfalls.
This survey provides dense coverage (every other night) and deep
imaging with its custom, broadband $VR$ filter.

\subsection{SuperMACHO and Supernova detection}
The SuperMACHO project is a five-year optical survey of the Large
Magellanic Cloud (LMC) aimed at detecting microlensing of LMC stars
(Stubbs et al. 2002).  The goal of this survey is to determine the
location of the lens population responsible for the excess
microlensing rate observed toward the LMC by the MACHO project (see
Alcock et al. 2000 and references therein) and, thereby, better
constrain the fraction of MAssive Compact Halo Objects (MACHOs) in the
Galactic halo. The survey is conducted on the CTIO Blanco 4m telescope
using a custom $VR$ broadband filter.  SuperMACHO observes 68 LMC
fields during dark and gray time in the months of October -- December.
We completed our fifth season of observations in the second half of
2005.  We process our images with a near-real time data reduction
pipeline that employs a difference-imaging technique (see Alard \&
Lupton 1998, Alcock et al. 1999, Alard 2000, and G{\"o}ssl 2002 )
which enables us to detect small changes in flux and to produce 
lightcurves uncontaminated by light from nearby, non-varying sources.

We present here a uniform set of densely sampled pre-maximum SNe~Ia
lightcurves from the SuperMACHO survey.  From these we constrain the
time to maximum brightness for SNe~Ia.  We present data to provide
constraints on SN~Ia explosion models to aid in discriminating between
competing theories.  In Section~\ref{section:obs} we discuss our
observations.  In Section~\ref{section:data} we present our data.  In
Section~\ref{section:discussion} we use our data to place limits on
the time to maximum brightness and present a functional model for the
SN~Ia lightcurve shape.

\section{Observations} \label{section:obs}

\subsection{Imaging}
\begin{figure}
\includegraphics[angle=0,width=8.4cm]{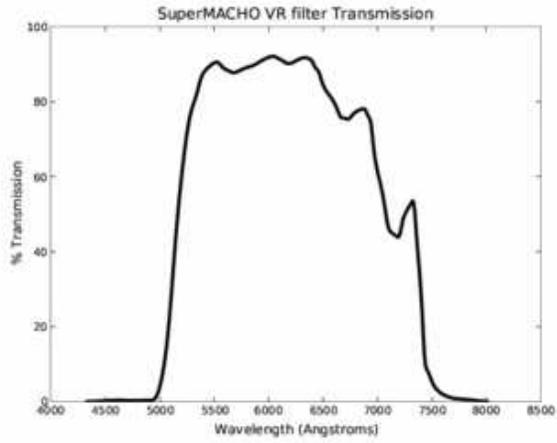}
\caption[$VR$ filter response]{Transmission curve for the SuperMACHO $VR$ filter.  An electronic table of the transmission curve can be found at http://ctiokw.ctio.noao.edu/$\sim$sm/sm/SNrise.}
\label{fig:vrresp}
\end{figure}

The lightcurves of the sources we report were obtained on the CTIO
Blanco 4m telescope during the 2004 season of the SuperMACHO survey.
The images were taken using the MOSAIC II wide-field CCD camera.  With
a plate scale of 0.27''/pixel, MOSAIC II's 8 SITe 2Kx4k CCDs cover a
0.32 sq. deg. field.  On a given night we image approximately 60 of
our 68 fields so that we obtain relatively dense time-coverage of the
events we detect.  All survey images are taken in a single, custom
$VR$ passband (see Figure~\ref{fig:vrresp} for transmission curve).
This broad filter enables us to detect flux excursions while they are
still too faint for many narrower filters to detect at high S/N.  We
use an Atmospheric Dispersion Corrector (ADC) to suppress the
atmospheric dispersion through our broad filter.  A detailed
description of the data reduction pipeline and event selection
criteria will be available in Rest et al. (2007, in preparation) and
Garg et al. (2007, in preparation).

The images are processed using a near real-time pipeline.  SuperMACHO
surveys 50 million sources.  The difference-imaging technique we use
enables us to limit our attention to a subset of those lightcurves
that includes only those that show changes in brightness.  We identify
candidate events by first choosing, from previous years' data, the
highest quality image for each field to create a set of templates.  We
then subtract the templates from the co-registered detection images to
produce ``difference images'' showing only sources whose brightness
has varied since the template epoch.  This difference-imaging
technique enables improved sensitivity to faint flux excursions,
particularly in crowded fields such as those in the LMC.  We consider
any difference flux detections coincident within a 1x1 pixel box in
all images of a field to be from a single source and so caused by a
unique flux excursion event. We obtain a difference lightcurve for
each flux excursion event by measuring the difference flux under a
point-source function whose center is forced to be at the centroid of
all the difference image detections clustered within that box.  By
performing this ``forced difference flux photometry'' on all images of
an event location, we measure changes in difference flux that are
below a triggering threshold of S/N $>$ 5.

Each night's data reveal hundreds of optically varying events.  The
majority of these are due to intrinsically variable stars, detector
artifacts, cosmic rays, and diffraction spikes from nearby bright
stars.  To limit the set of lightcurves to unique flux excursions
(such as microlensing, AGN activity, and supernovae) of real sources,
a series of cuts are applied to the lightcurves.  These include the
significance of the measured difference flux and goodness-of-fit to a
flat baseline in years prior to the event.  Known variable sources in
the MACHO catalog and sources with more than 3 difference detections
of S/N $>$ 10 in previous years are removed from the set.  Finally,
all remaining lightcurves and their associated detection and
difference images are inspected by eye to remove spurious detections
caused by artifacts.  This selection process whittles the set of new
candidate transient events discovered each night of the survey to
approximately 20.  Fits to models of microlensing and SN~Ia
lightcurves and visual inspection of template and difference images
(for the appearance of host galaxies) are used to preliminarily
categorize the events as microlensing, supernovae, AGN's, or other
optical transients.  The events are then placed in a queue for
spectroscopic confirmation (see Section~\ref{section:specobs}).

The final lightcurves we present in this paper were produced using the
N(N-1)/2 method (hereafter ``NN2'') of Barris et al. (2005).  With
this method, instead of using a single template, we difference all
possible image pairs to produce the final lightcurve whose points
are weighted combinations of the difference flux in all subtractions
for a given observation.  We use NN2 subtractions to provide cleaner
difference lightcurves for our SNe which are behind very crowded LMC
fields and often close to other variable sources.

\subsection{Spectroscopy} \label{section:specobs}

Both Magellan telescopes, Clay and Baade, were used to to obtain
spectroscopic follow-up of events identified by the CTIO 4m.  On the
Clay Telescope, the Low Dispersion Survey Spectrograph 2 (LDSS2;
Allington-Smith et al., 1994) was used to obtain longslit spectroscopy
on our targets.  The LDSS2 CCD detector has a resolution of
0.378"/pixel. We used the following configuration for the spectra
obtained on this instrument: the medium resolution (300~l/mm) blue
grism blazed at 5000\AA, a slit of 0.75", and no blocking filter.  The
spectra have a nominal dispersion of 5.3\AA/pix over the useful
wavelength range of $\sim$3800--7500\AA.  On the Baade Telescope, we
used the Inamori-Magellan Areal Camera and Spectrograph (IMACS;
Bigelow \& Dressler, 2003) in longslit mode with the long camera (f/4
focus) and the medium resolution, 300~l/mm, grating.  In this
configuration the instrument provides a 0.111"/pix image scale with a
nominal dispersion of 0.743\AA/pix over a useful wavelength span of
3800--7500\AA\ without order blocking filters.  The nights were mostly
photometric and the Shack-Hartmann wavefront sensor provided image
qualities of $\sim$0.6"--1.1" FWHM.  To minimize slit losses due to
atmospheric dispersion, we used a slit aligned to the parallactic
angle.  Observations typically consisted of multiple integrations on a
source.  The S/N on each target varied with the integration times,
source brightness, transparency, and seeing.

Reduction of the spectra consists of the typical single slit
processing using standard IRAF routines for bias subtraction and
flat-fielding.  Cosmic ray removal is facilitated using the Laplacian
Cosmic Ray identification routine of van Dokkum (2001).  We co-add the
processed 2D images of each target and extract 1D apertures using
isolated regions around the target source for the background
subtraction.  We determine the best 1D extraction by iterating through
multiple target and sky regions to ensure proper source and sky
isolation within the crowded LMC fields. We find the dispersion
solution for each image using He~Ne~Ar arc lamp observations that show
a typical RMS of $<$0.5\AA.  We use spectrophotometric standards
(Feige 110, Hiltner 600, and LTT3864) observed on the same night as
the targets for flux calibratation.

\section{Data} \label{section:data}

\subsection{Lightcurves}
\begin{figure}
\includegraphics[angle=0,width=8.4cm]{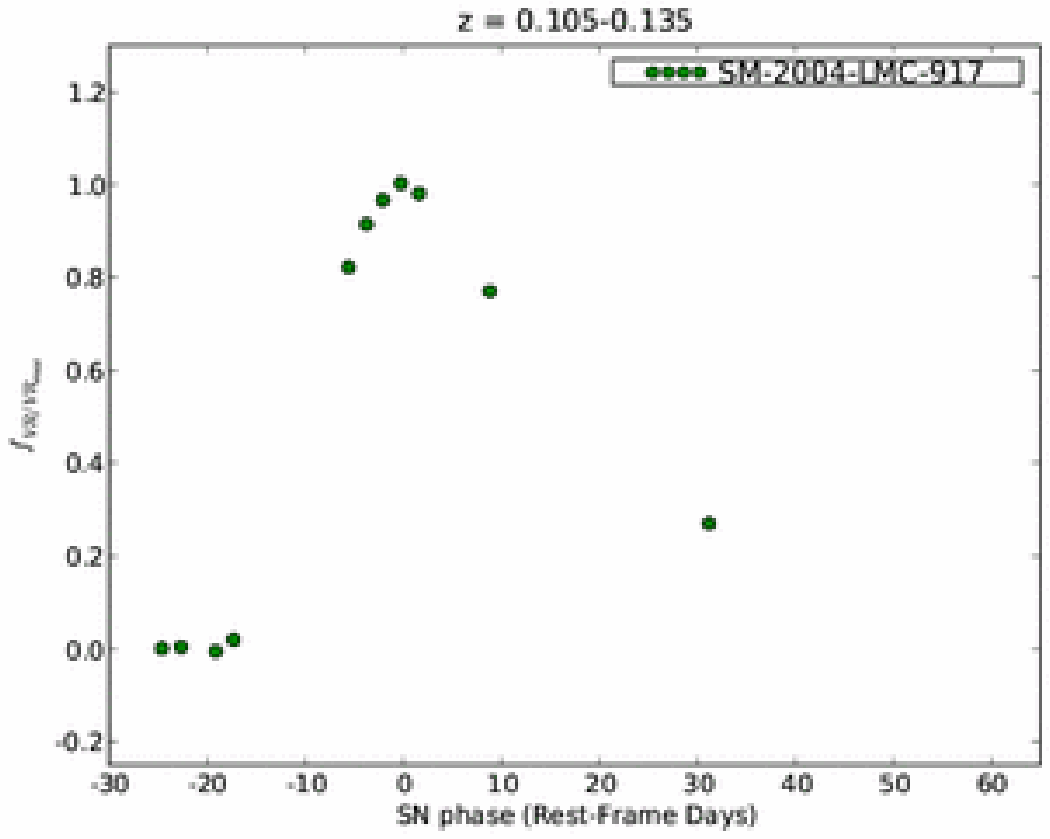}
\caption[Lightcurves of z = 0.105--0.135]{Restframe $VR$-band lightcurve of
SM-2004-LMC-917 which has a redshift of 0.11.  Data are shown in flux units normalized to the flux at maximum brightness.  Error bars represent $1\sigma$ errors.
Where error bars are not seen, $1\sigma$ errors are smaller than
symbol. }
\label{fig:z11}
\end{figure}

\begin{figure}
\includegraphics[angle=0,width=8.4cm]{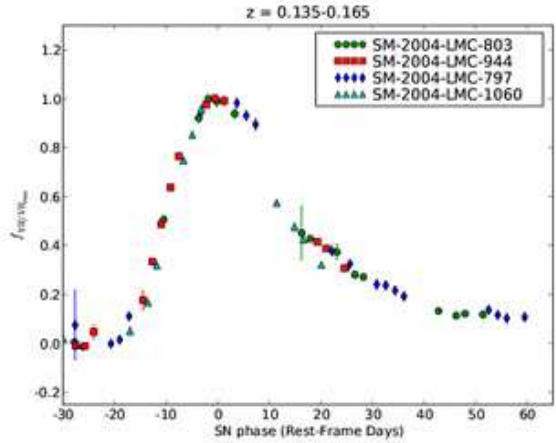}
\caption[Lightcurves of z = 0.135--0.165]{Restframe $VR$ lightcurves of four
SNe~Ia at z = 0.135--0.165. Data are shown in flux units normalized to the flux at maximum brightness.  Error bars represent $1\sigma$ errors.
Where error bars are not seen, $1\sigma$ errors are smaller than
symbol.  SM-2004-LMC-803 (circles) has a redshift of 0.16.
SM-2004-LMC-944 (squares) has a redshift of 0.15. SM-2004-LMC-797
(diamonds) has a redshift of 0.145.  SM-2004-LMC-1060 (triangles) has a
 redshift of 0.154.}
\label{fig:z15}
\end{figure}

\begin{figure}
\includegraphics[angle=0,width=8.4cm]{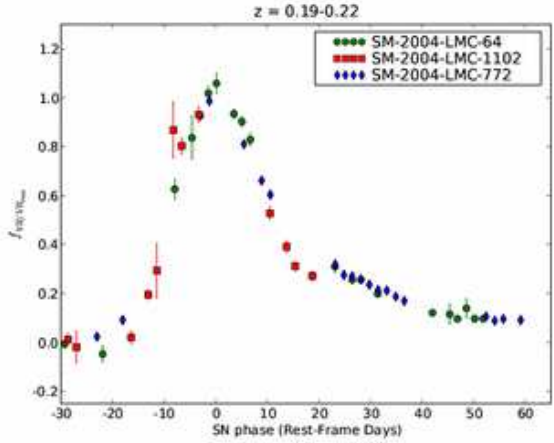}
\caption[Lightcurves of z = 0.19--0.22]{Restframe $VR$ lightcurves of three
SNe~Ia at z = 0.19--0.22.  Data are shown in flux units normalized to the flux at maximum brightness.  Error bars represent $1\sigma$ errors.
Where error bars are not seen, $1\sigma$ errors are smaller than
symbol.  SM-2004-LMC-64 (circles) has a redshift of
0.22.  SM-2004-LMC-1102 (squares) has a redshift of
0.22. SM-2004-LMC-772 (diamonds) has a redshift of 0.19.}
\label{fig:z20}
\end{figure}

\begin{figure}
\includegraphics[angle=0,width=8.4cm]{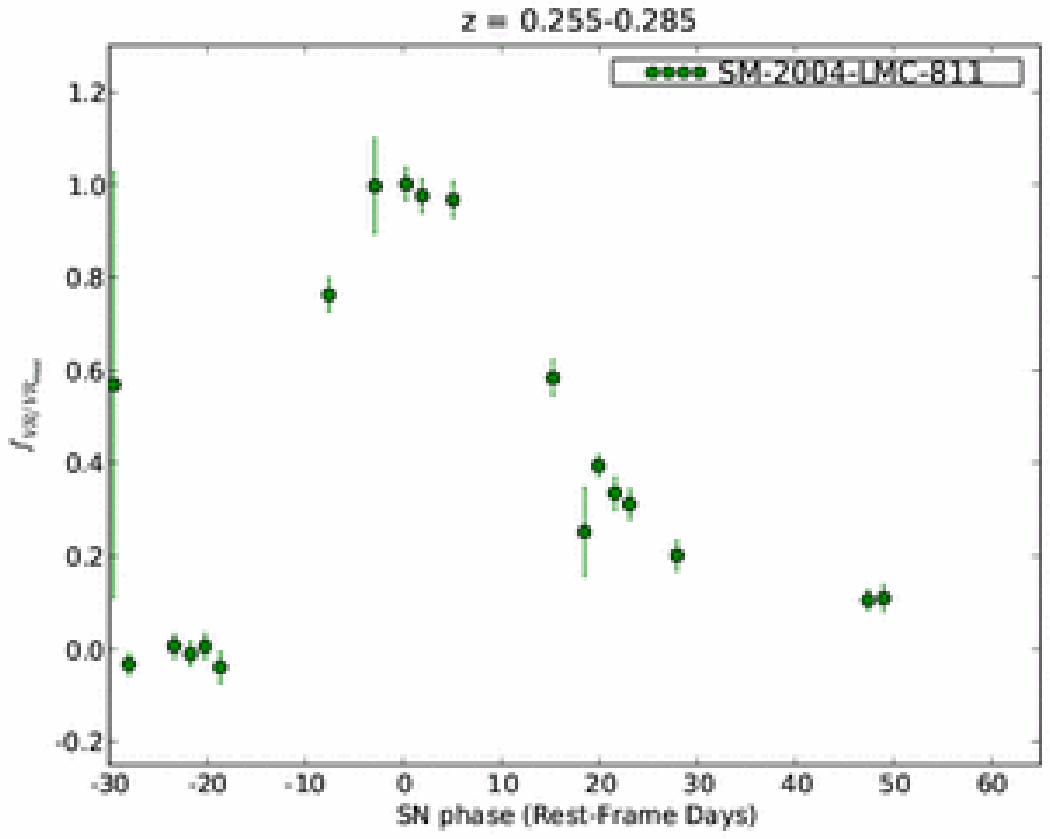}
\caption[Lightcurves of z = 0.255--0.285]{Restframe $VR$ lightcurve of SM-2004-LMC-811 which has a redshift of 0.27. Data are shown in flux units normalized to the flux at maximum brightness.  Error bars represent $1\sigma$ errors.
Where error bars are not seen, $1\sigma$ errors are smaller than
symbol. }
\label{fig:z27}
\end{figure}

\begin{figure}
\includegraphics[angle=0,width=8.4cm]{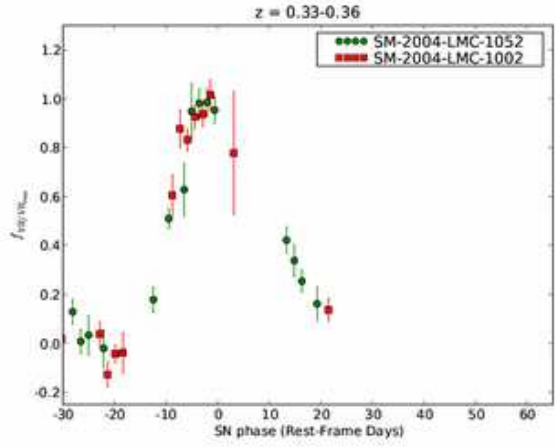}
\caption[Lightcurves of z = 0.33--0.36]{Restframe $VR$ lightcurves of two
  SNe~Ia at z = 0.33--0.36.  Data are shown in flux units normalized to the flux at maximum brightness.  Error bars represent $1\sigma$ errors.
  Where error bars are not seen, $1\sigma$ errors are smaller than
  symbol.  SM-2004-LMC-1052 (circles) has a redshift of 0.348.
  SM-2004-LMC-1002 (squares) has a redshift of 0.35.}
\label{fig:z34}
\end{figure}

We present 11 SNe~Ia from the 2004 observing season.
Table~\ref{tab:SNinfo} gives their positions and redshifts.
Tables~\ref{tab:2004-LMC-64.tab}--\ref{tab:2004-LMC-1102.tab} give the
lightcurves for each object.  The NN2 difference fluxes in the
lightcurves are given normalized to a zero point of 25 (see
Rest~et~al.~2005 for $VR$-band standardization procedure).
Figures~\ref{fig:z11}--\ref{fig:z34} show the lightcurves with the
time axis transformed to the SN rest-frame and relative to the time of
maximum brightness in the observed $VR$-band, $t_{max}$ (see
Section~\ref{section:tmax} for $t_{max}$ determination procedure).  We
normalize the observed fluxes to the flux at maximum, $VR_{max}$ to
obtain the $f_{\frac{VR}{VR_{max}}}$ lightcurves shown.  The SNe are
grouped by redshift, and each figure shows all SNe with similar
redshifts (see Section~\ref{section:Spectra} for redshift
determination procedure).  We group the SNe~Ia by redshift to limit
the impact of K-corrections (Hamuy et al. 1993, Kim et al. 1996,
Schmidt et al. 1998, and Nugent et al. 2002) on our findings (see
\ref{section:kcorr} for further discussion of K-corrections).

\subsection{Spectra} \label{section:Spectra}
\begin{figure}
\includegraphics[angle=0,width=8.4cm]{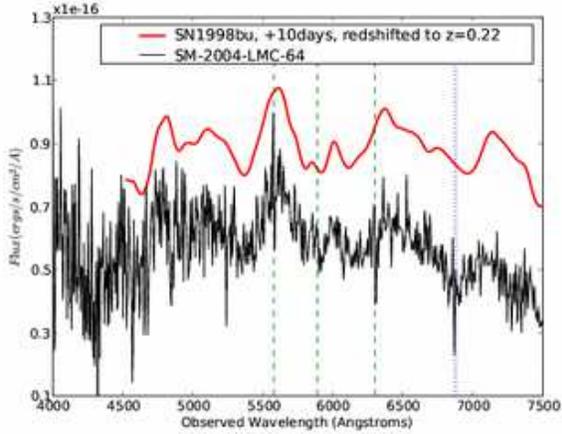}
\caption[Spectrum of SM-2004-LMC-64]{Flux-calibrated spectrum of SM-2004-LMC-64 with comparison nearby spectrum of SN1998bu above.  The spectrum of SN1998bu was taken at +10 days relative to $B$-band maximum and is shown redshifted to z = 0.22.  The flux of the comparison spectrum has been smoothed, scaled, and offset.  Dashed lines indicate sky emission lines at 5577\AA, 5890\AA, and 6301\AA.  Dotted lines demark the atmospheric O$_{2}$-band between 6867--6884\AA.  Electronic data tables can be found at http://ctiokw.ctio.noao.edu/$\sim$sm/sm/SNrise.}
\label{fig:lmc64}
\end{figure}

\begin{figure}
\includegraphics[angle=0,width=8.4cm]{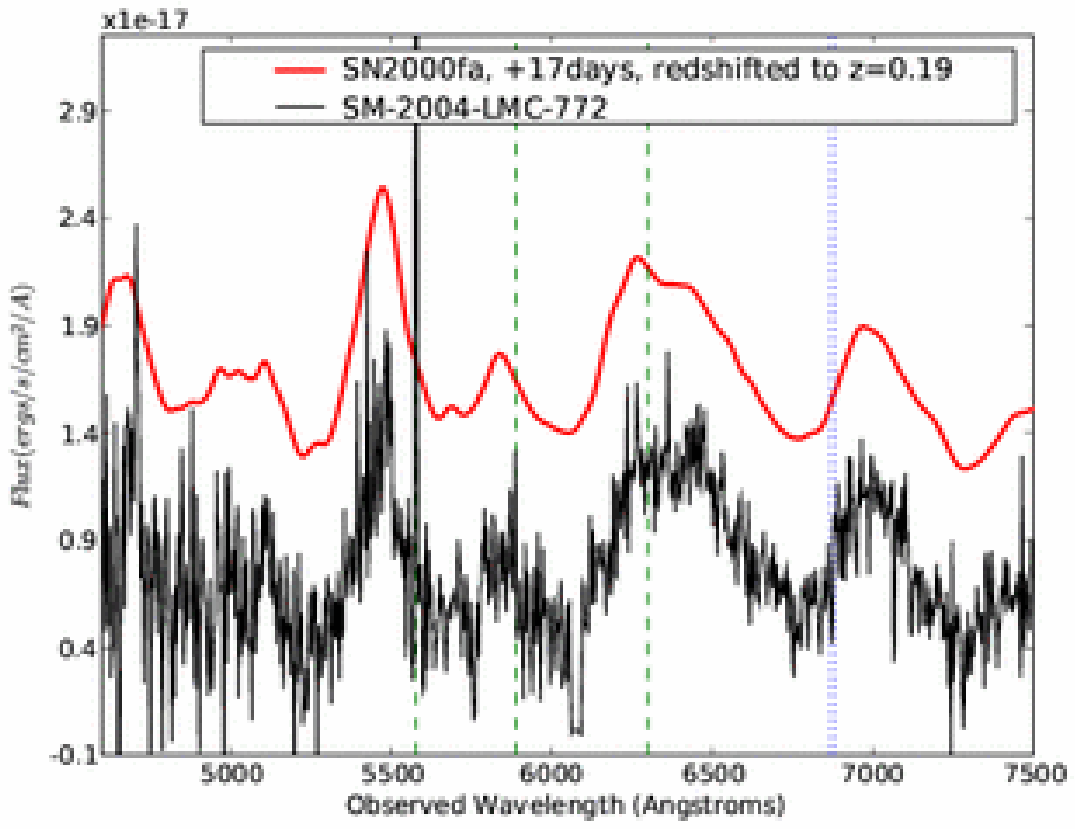}
\caption[Spectrum of SM-2004-LMC-772]{Flux-calibrated spectrum of SM-2004-LMC-772 with comparison nearby spectrum of SN2000fa above.  The spectrum of SN2000fa was taken at +17 days relative to $B$-band maximum and is shown redshifted to z = 0.19.  The flux of the comparison spectrum has been smoothed, scaled, and offset.  See Figure~\ref{fig:lmc64} for explanation of dashed and dotted lines.}
\label{fig:lmc772}
\end{figure}

\begin{figure}
\includegraphics[angle=0,width=8.4cm]{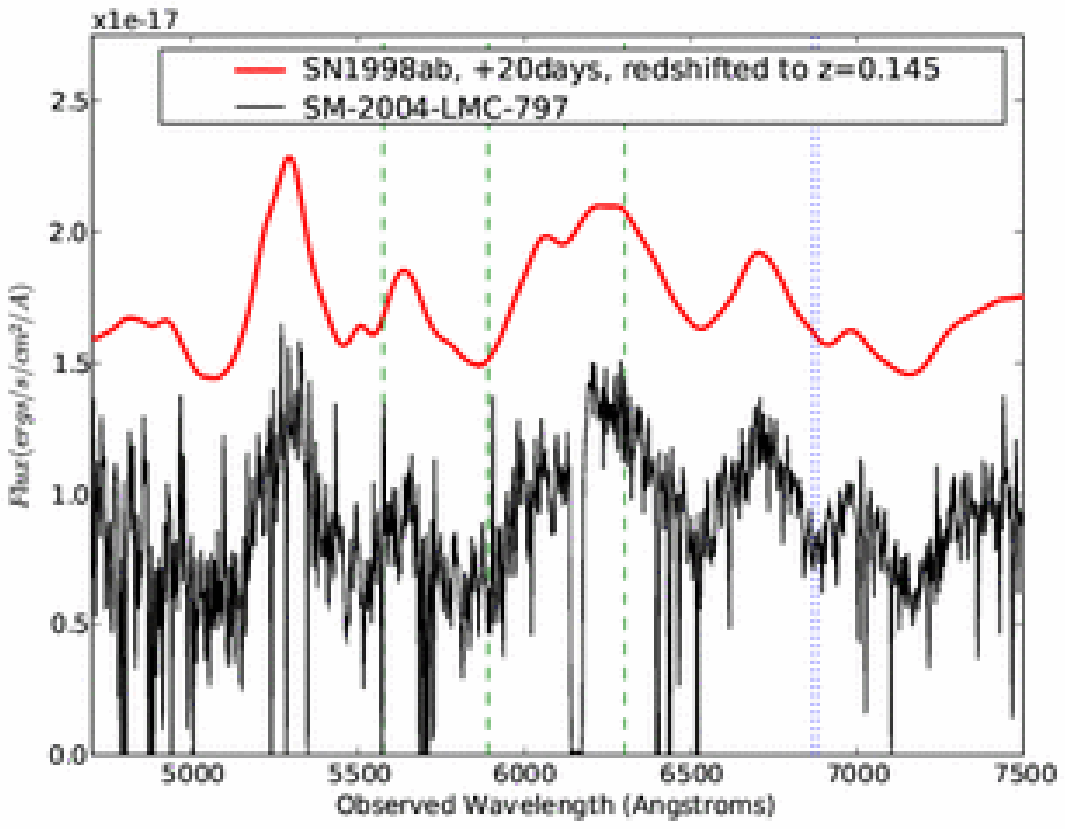}
\caption[Spectrum of SM-2004-LMC-797]{Flux-calibrated spectrum of SM-2004-LMC-797 with comparison nearby spectrum of SN1998ab above.  The spectrum of SN1998ab was taken at +20 days relative to $B$-band maximum and is shown redshifted to z = 0.145.  The flux of the comparison spectrum has been smoothed, scaled, and offset.  See Figure~\ref{fig:lmc64} for explanation of dashed and dotted lines.}
\label{fig:lmc797}
\end{figure}

\begin{figure}
\includegraphics[angle=0,width=8.4cm]{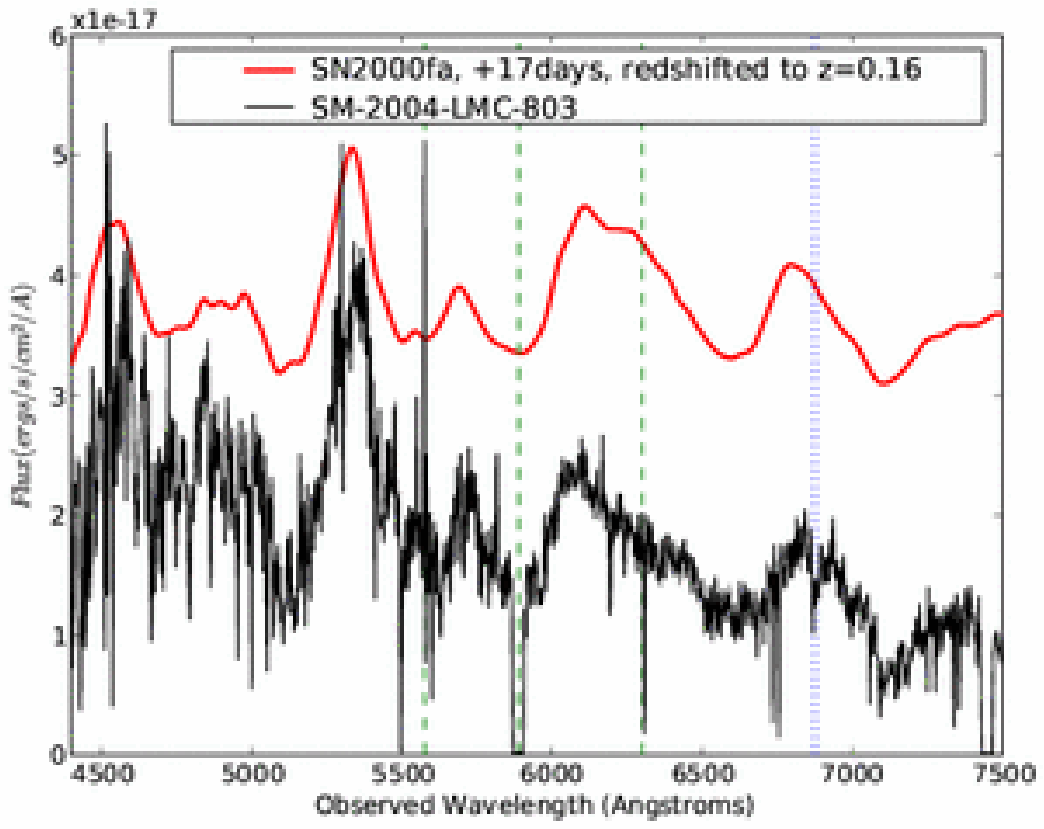}
\caption[Spectrum of SM-2004-LMC-803]{Flux-calibrated spectrum of SM-2004-LMC-803 with comparison nearby spectrum of SN2000fa above.  The spectrum of SN2000fa was taken at +17 days relative to $B$-band maximum and is shown redshifted to z = 0.16.  The flux of the comparison spectrum has been smoothed, scaled, and offset.  See Figure~\ref{fig:lmc64} for explanation of dashed and dotted lines.}
\label{fig:lmc803}
\end{figure}

\begin{figure}
\includegraphics[angle=0,width=8.4cm]{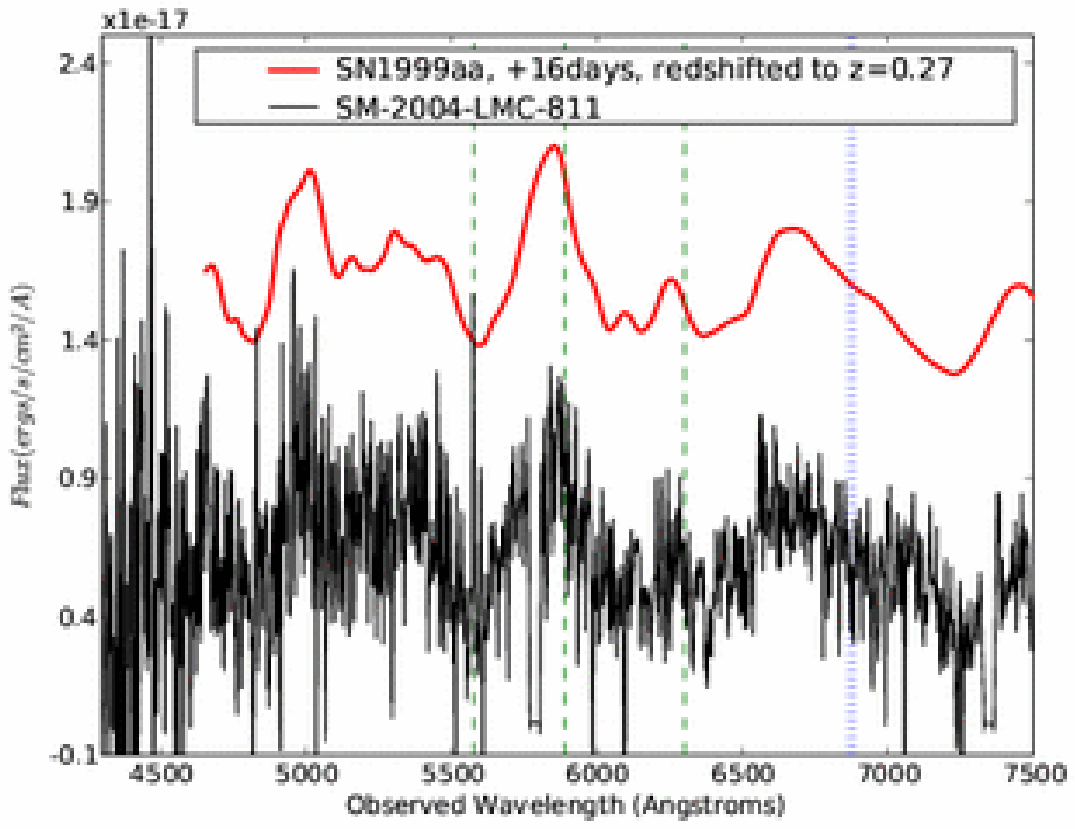}
\caption[Spectrum of SM-2004-LMC-811]{Flux-calibrated spectrum of SM-2004-LMC-811 with comparison nearby spectrum of SN1999aa above.  The spectrum of SN1999aa was taken at +16 days relative to $B$-band maximum and is shown redshifted to z = 0.27.  The flux of the comparison spectrum has been smoothed, scaled, and offset.  See Figure~\ref{fig:lmc64} for explanation of dashed and dotted lines.}
\label{fig:lmc811}
\end{figure}

\begin{figure}
\includegraphics[angle=0,width=8.4cm]{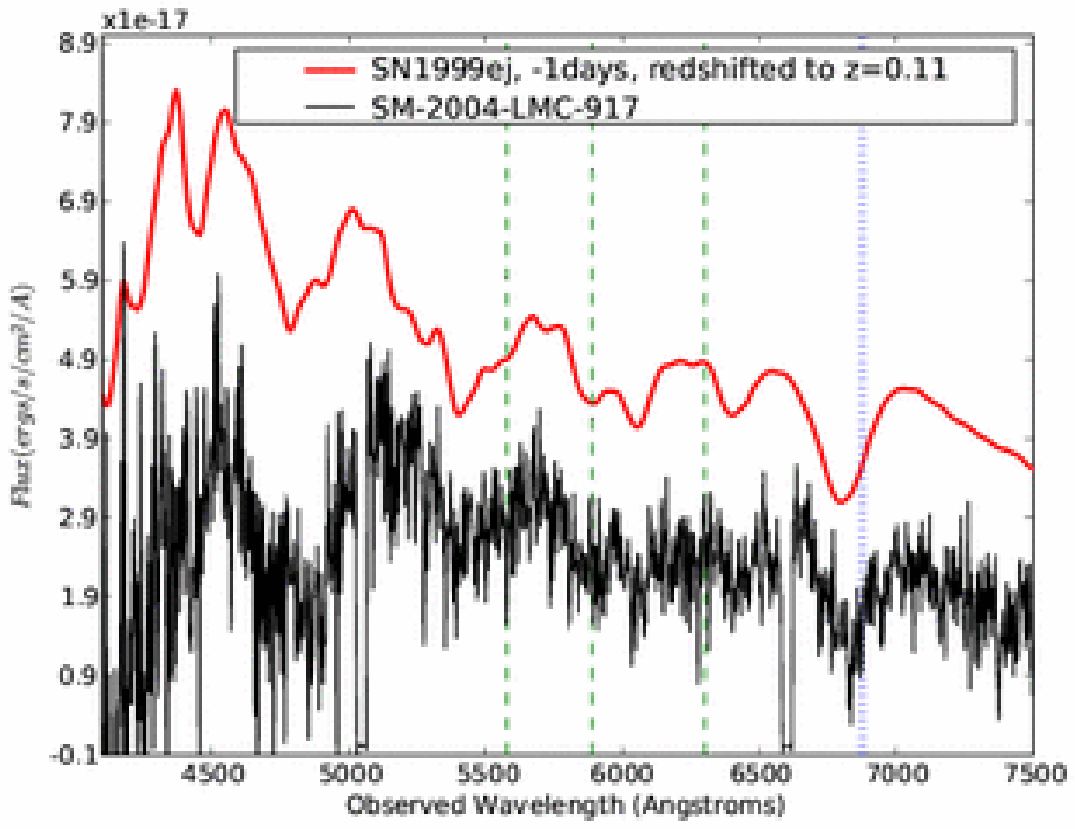}
\caption[Spectrum of SM-2004-LMC-917]{Flux-calibrated spectrum of SM-2004-LMC-917 with comparison nearby spectrum of SN1999ej above.  The spectrum of SN1999ej was taken at -1 days relative to $B$-band maximum and is shown redshifted to z = 0.11.  The flux of the comparison spectrum has been smoothed, scaled, and offset.  See Figure~\ref{fig:lmc64} for explanation of dashed and dotted lines.}
\label{fig:lmc917}
\end{figure}

\begin{figure}
\includegraphics[angle=0,width=8.4cm]{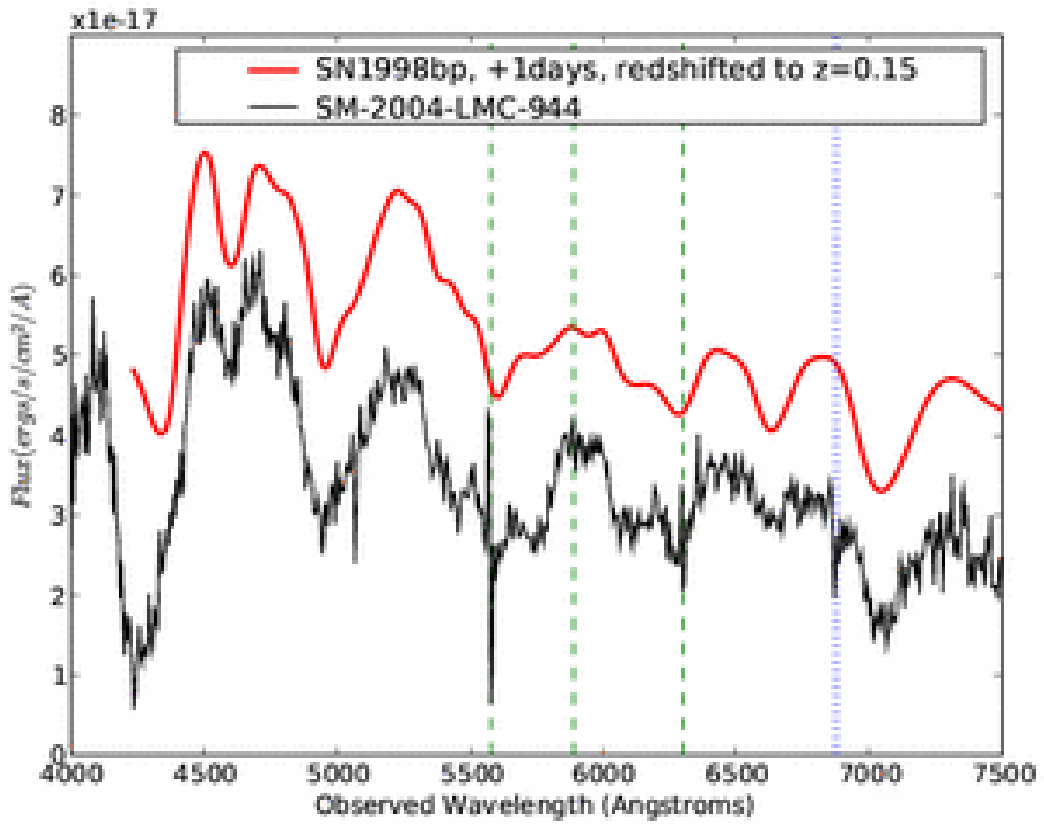}
\caption[Spectrum of SM-2004-LMC-944]{Flux-calibrated spectrum of SM-2004-LMC-944 with comparison nearby spectrum of SN1998bp above.  The spectrum of SN1998bp was taken at +1 days relative to $B$-band maximum and is shown redshifted to z = 0.15.  The flux of the comparison spectrum has been smoothed, scaled, and offset.  See Figure~\ref{fig:lmc64} for explanation of dashed and dotted lines.}
\label{fig:lmc944}
\end{figure}

\begin{figure}
\includegraphics[angle=0,width=8.4cm]{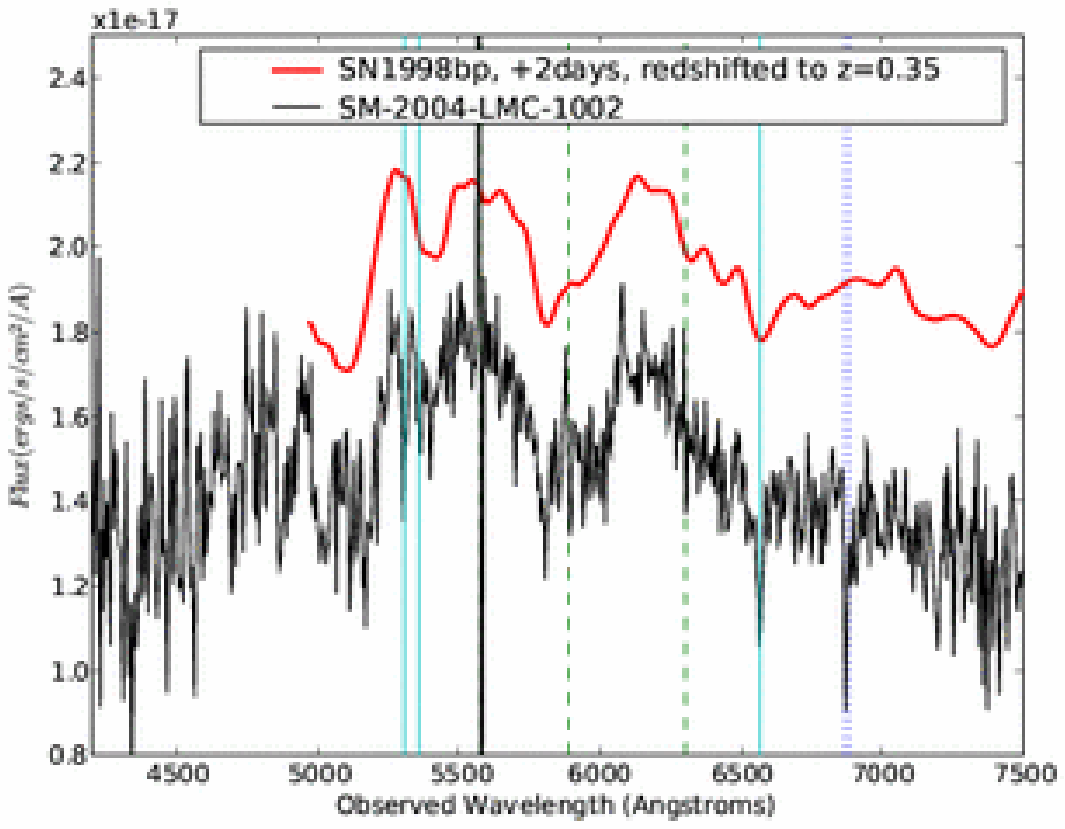}
\caption[Spectrum of SM-2004-LMC-1002]{Flux-calibrated spectrum of SM-2004-LMC-1002 with comparison nearby spectrum of SN1998bp above.  The spectrum of SN1998bp was taken at +2 days relative to $B$-band maximum and is shown redshifted to z = 0.35.  The flux of the comparison spectrum has been smoothed, scaled, and offset.  Solid lines mark galaxy features used to independently find the source redshift of z = 0.350.  See Figure~\ref{fig:lmc64} for explanation of dashed and dotted lines.}
\label{fig:lmc1002}
\end{figure}

\begin{figure}
\includegraphics[angle=0,width=8.4cm]{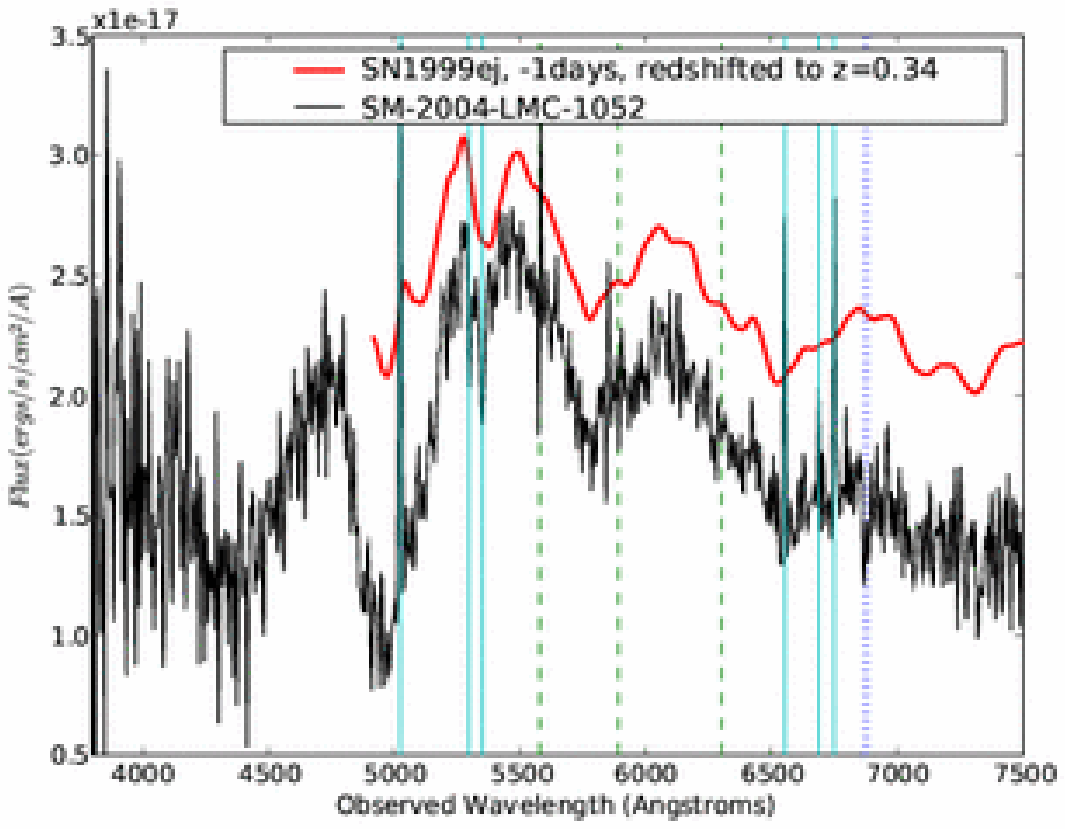}
\caption[Spectrum of SM-2004-LMC-1052]{Flux-calibrated spectrum of SM-2004-LMC-1052 with comparison nearby spectrum of SN1999ej above.  The spectrum of SN1999ej was taken at -1 days relative to $B$-band maximum and is shown redshifted to z = 0.34.  The flux of the comparison spectrum has been smoothed, scaled, and offset.  Solid lines mark galaxy features used to independently find the source redshift of z = 0.348.  See Figure~\ref{fig:lmc64} for explanation of dashed and dotted lines.}
\label{fig:lmc1052}
\end{figure}

\begin{figure}
\includegraphics[angle=0,width=8.4cm]{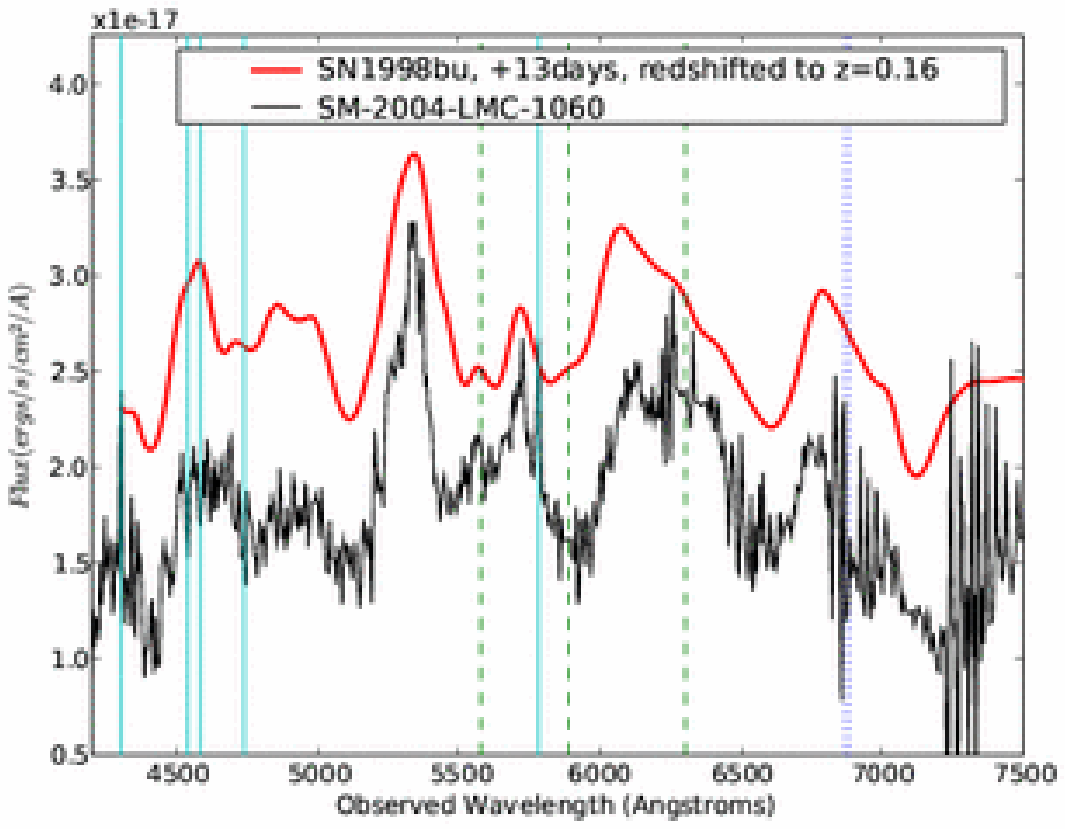}
\caption[Spectrum of SM-2004-LMC-1060]{Flux-calibrated spectrum of SM-2004-LMC-1060 with comparison nearby spectrum of SN1998bu above.  The spectrum of SN1998bu was taken at +13 days relative to $B$-band maximum and is shown redshifted to z = 0.16.  The flux of the comparison spectrum has been smoothed, scaled, and offset.  Solid lines mark galaxy features used to independently find the source redshift of z = 0.154.  See Figure~\ref{fig:lmc64} for explanation of dashed and dotted lines.}
\label{fig:lmc1060}
\end{figure}

\begin{figure}
\includegraphics[angle=0,width=8.4cm]{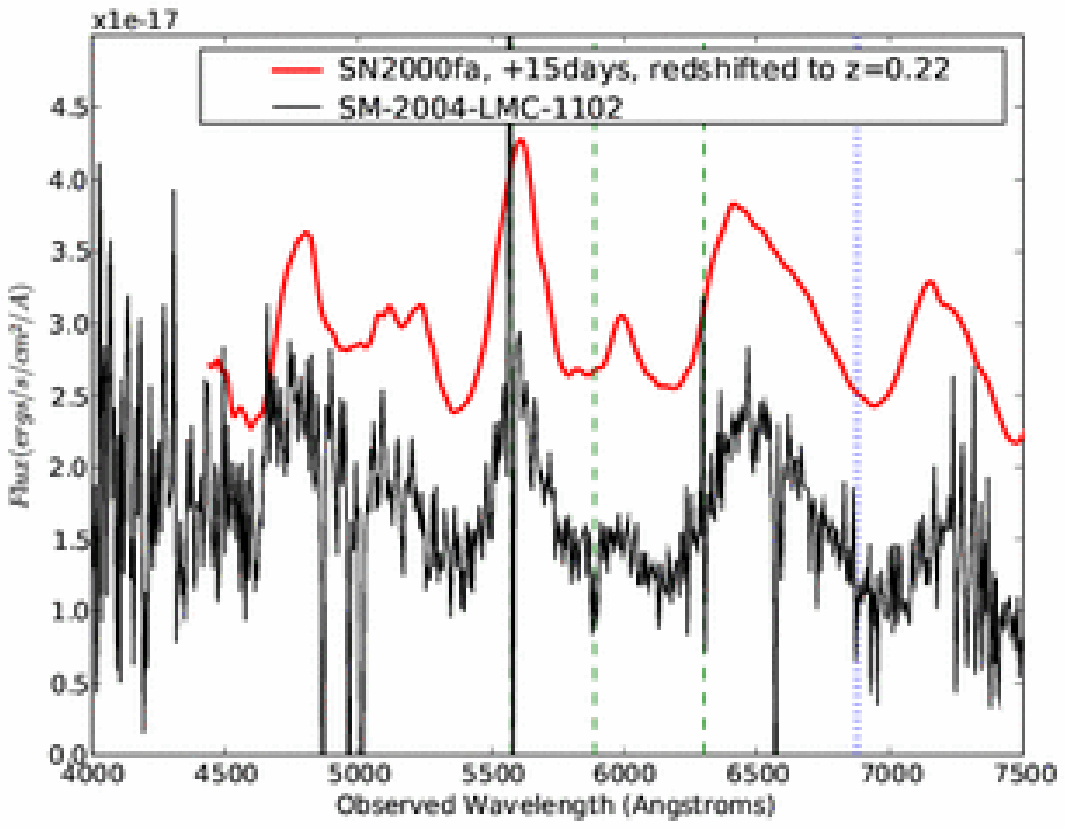}
\caption[Spectrum of SM-2004-LMC-1102]{Flux-calibrated spectrum of SM-2004-LMC-1102 with comparison nearby spectrum of SN2000fa above.  The spectrum of SN2000fa was taken at +15 days relative to $B$-band maximum and is shown redshifted to z = 0.22.  The flux of the comparison spectrum has been smoothed, scaled, and offset.  See Figure~\ref{fig:lmc64} for explanation of dashed and dotted lines.}
\label{fig:lmc1102}
\end{figure}

Table~\ref{tab:specinfo} lists the telescope, instrument, observation
date, and total integration time for each spectrum presented.  We
determine the SN type and redshift by comparing the spectrum to a
library of nearby SN spectra (Matheson et al. 2006, in preparation).
Following the method of Matheson et al. (2005) we classify an event as
a SN~Ia if it shows the characteristic CaII H\&K, SiII, FeII, and SII
features (Filippenko 1997).  We choose a comparison spectrum from the
nearby library that was obtained at approximately the same SN phase as
our spectrum. We determine the object's redshift by redshifting the
nearby spectrum until the peaks and valleys match.  This gives z to an
accuracy of $\sim$0.01.  Because we do not apply Galactic, LMC, or
host galaxy reddening corrections, the continuum shapes of our spectra
sometimes appear flatter and redder than that of the nearby,
reddening-corrected spectrum.
Figures~\ref{fig:lmc64}--\ref{fig:lmc1102} show the spectrum of each
SN with the redshifted nearby comparison spectrum above.  The SN's
redshift determined by this method is given in
Table~\ref{tab:SNinfo}.

Three of the spectra also exhibit strong host galaxy emission and
absorption features.  We use these features to obtain more accurate
redshifts for these sources and to verify the nearby SN comparison
method of redshift determination used for the remaining SNe.  To
determine the galactic redshifts we first find the line centers of the
emission and absorption features by fitting a Gaussian profile to
each.  We then calculate the galaxy's redshift by averaging the
redshifts of the identified lines.  Table~\ref{tab:galinfo} lists the
SNe whose spectra exhibit strong galaxy features, the lines seen, and
the galaxy redshift.  For reference, the table also lists the redshift
determined by the nearby comparison method.  In all cases, the
redshifts found by the two methods agree within better than 0.01.

\section{Discussion} \label{section:discussion}

\subsection{Functional Model of SN~Ia Lightcurve} \label{section:templateLC}

To model our observed $VR$-band lightcurves, we choose the following
function, $\phi(t)$, to describe the difference flux normalized to the
difference flux at the time of maximum brightness in the $VR$-band:

\noindent $\phi  = 0.0$ for $t < t_{r}$\\
$\phi = \frac{(t-t_{r})^{2}}{t_{r} (t_{r} - n)}$ for $t_{r} < t < n$\\
$\phi = 1 - \frac{t^{2}}{n t_{r}}$ for $n < t < 0$\\
$\phi = 1 - \gamma t^{2}$ for $0 < t < m$\\
$\phi = 1 - m^{2} \gamma + 2m \gamma \tau (e^{\frac{m-t}{\tau}} - 1)$ for $t > m$\\

\noindent where $t$ is the SN phase in rest-frame days scaled such
that $t = 0$ is the time of maximum, $\phi$ is the ratio of observed
$VR$-band flux at $t$ to maximum flux, $t_{r}$ is the time of
explosion, $n$ and $m$ are arbitrary SN phases such that $n < 0$ and
$m > 0$, $\gamma$ is an arbitrary constant, and $\tau$ measures the
decay time of the late-time lightcurve.  The early-time portion of our
model is motivated by Riess~et~al.~(1999).  Riess et al.~fit their
SN~Ia lightcurves prior to $-10$ days with the expanding fireball
model of Goldhaber et al. (1998) which has the functional form of a
parabola with a minimum at the time of explosion.  We model the
expanding fireball as $\phi~=~\alpha (t-t_{r})^{2}$.  We choose an
exponential for the late-time lightcurve shape because we expect the
luminosity to be dominated by radioactive decay.  For the exponential
we pick the generic form
$\phi~=~\phi_{o}e^{-\frac{t-t_{m}}{\tau}}~+~c$.  The form of $\phi$
between $-n < t < 0$ and $0 < t < m$ is taken to be two arbitrary
second degree polynomials constrained to be 1 at $t$ = 0.  We use the
forms $\phi~=~1~-~\beta t^{2}$ and $\phi~=~1~-~\gamma t^{2}$
respectively.  We leave $n$ and $m$ as free parameters in our fit.  By
requiring that $\phi$ be a smoothly connected function (i.e. that the
value of $\phi$ and its first derivative are everywhere continous), we
eliminate $\alpha$, $\beta$, $c$, $\phi_{o}$, and $t_{m}$.  This
results in the form of $\phi$ given above, with $t_{r}$, $\tau$, $n$,
$m$, and $\gamma$ as the 5 remaining free parameters.

In the following sections, we will use this model to estimate the
time, $t_{max}$, of observed frame maximum brightness, $VR_{max}$, for
each SNe and to place constraints on the interval between the
time-of-explosion and maximum brightness.

\subsection{Estimation of $t_{max}$} \label{section:tmax}
For each SN presented, we determine $t_{max}$ and $VR_{max}$ using the
functional SN~Ia model presented in Section~\ref{section:templateLC}.
We do so by adding $t_{max}$ and $VR_{max}$ as free parameters to the
model such that

\noindent $f_{obs}(t_{obs}) = VR_{max} \phi(\frac{t_{obs}-t_{max}}{z})$

\noindent where $f_{obs}$ is the observed flux, $t_{obs}$ is the
time of the observation, and $z$ is the SN redshift.

Using the C-MINUIT implementation of the MINUIT\footnote{See
http://wwwasdoc.web.cern.ch/wwwasdoc/minuit/minmain.html for
documentation on the MINUIT package.} minimization package, we
individually determine the best fit for each lightcurve to $f_{obs}$
by minimizing $\chi^{2}$.  Table~\ref{tab:SNinfo} gives the $t_{max}$
and $VR_{max}$ values for each SN along with the parabolic errors
returned by the MIGRAD processor in MINUIT.  We emphasize that these
fits are performed on the lightcurves as observed with no
K-corrections, reddening corrections, or adjustments to account for
SN~Ia lightcurve shape.  We use these fits to obtain estimates of
$t_{max}$ and $VR_{max}$ for each SN and not to assess whether our
model, $\phi(t)$, provides a good description of the SN~Ia lightcurve.
We will discuss the validity of our model below in
Section~\ref{section:compositeLC}.  For now we choose this model to
estimate $t_{max}$ and $VR_{max}$ because we assume that the SN~Ia
lightcurve is a smooth, continuous function with single maximum and an
asymmetric shape.  $f_{obs}(t_{obs})$ provides a generic model for
such a curve and should give a reasonable description of the maximum.
To provide an initial assessment of this assumption we note that for
each SN the best fit curve generally has a $\chi^{2}$ value close to
1.

For each SN, we use our estimation of $t_{max}$ and its measured
redshift to determine the rest-frame phase, relative to $t_{max}$, of
each observation.  In the cases where a galaxy redshift is available
(see Table~\ref{tab:galinfo}), we use its value for the SN's redshift.
Tables~\ref{tab:2004-LMC-64.tab}--\ref{tab:2004-LMC-1102.tab} give the
phase and the significance (S/N) of each measurement.
Table~\ref{tab:SNinfo} lists the phase of the first S/N$>$5
observation for each SN.

We scale the difference fluxes to $VR_{max}$ and correct for time
dilation using the redshift determined in
Section~\ref{section:Spectra} to obtain the lightcurves shown in
Figures~\ref{fig:z11}--\ref{fig:z34}.  The SNe are presented grouped
by redshift to minimize the differences in the K-corrections for the
SNe in each group.  We expect the observed frame $VR$-band lightcurve
to vary with redshift as the $VR$ filter samples different portions of
the rest-frame spectrum.  Because the spectra of SNe~Ia near the time
of explosion are not well-studied and because we lack multi-epoch
multi-band data for our lightcurves, we bin our data by redshift
rather than apply K-corrections.  We choose a bin size of 0.03 in
redshift to maximize the number of SNe per bin while keeping the
difference in K-corrections between redshifts within a bin small.


\subsection{Construction of Composite of SN~Ia Lightcurve} \label{section:compositeLC}
Using the normalized $f_{\frac{VR}{VR_{max}}}$ lightcurves presented in
Section~\ref{section:tmax}, we construct a composite SN~Ia lightcurve
that is well-sampled from the time of explosion to $+$60 days.  We
include SNe from the redshift bin z=0.135--0.165 to create the
composite.  For this redshift bin the center of our broadband filter
corresponds to approximately 5200\AA\ in the rest-frame, close to
$V$-band.  We would expect the light passing through this filter to be
continuum-dominated, though some FeII \& III, SiII, and SII features
are present (Filippenko 1997).  We use our composite lightcurve to
examine the SN~Ia lightcurve.  In particular we discuss how well the
functional form presented in Section~\ref{section:templateLC}
describes the lightcurve shape by performing a multi-parameter fit to
the composite lightcurve.  We also discuss the rise time to maximum
brightness as parameterized by $t_{r}$ in our functional model.

Using C-MINUIT to minimize $\chi^{2}$, we perform a multi-parameter
fit of $\phi(t)$ to the composite lightcurve, including only data
between $-30$ rest-frame days and $+60$ rest-frame days so as not to
allow the flat baseline to dominate the $\chi^{2}$ of our best fit.
Though we fit all four SNe simultaneously, we also refit for $t_{max}$
and $VR_{max}$ of each individual SN in the composite.  For each SN,
the best fit $t_{max}$ obtained through the simultaneous fit agrees
with the $t_{max}$ found in the individual fits in
Section~\ref{section:tmax} to within one observed-frame day.  An
initial fit to all four SNe in the z=0.135--0.165 bin indicates that
SM-2004-LMC-1060 is a much faster decliner than the other SNe in the
bin, a result that can be verified from a qualitative inspection of
Figure~\ref{fig:z15}.  Removing this SN from the composite lightcurve,
we refit $\phi(t)$ and find a best fit $\chi^{2}/d.o.f.$ of 1.16 for
38 $d.o.f.$.  A summary of the parameters and their $1\sigma$
parabolic error uncertainties is given in Table~\ref{tab:parameters}.

From this fit we conclude that our functional model provides a
reasonable description of the overall shape of the observed $VR$-band
lightcurve for a SN~Ia with z between 0.135 and 0.165.  To draw
further conclusions about the SN~Ia lightcurve from the best fit
parameters, we must discuss them in the context of the systematic
effects that might alter the overall composite lightcurve shape and
also of any effects introduced by using multiple SNe with different
systematics to create the composite.  We discuss the three largest
systematic effects affecting our composite lightcurve: 1) the lack of
K-corrections to account for SNe at different redshifts; 2) intrinsic
diversity in the SN~Ia family; and 3) reddening from the host
galaxies, the LMC, and the Milky Way.

To examine the effects of the systematics, we create a tool to
construct empirical models of observed $VR$-band lightcurves using a
library of nearby SN~Ia spectra and lightcurves.  The lightcurve
library spans a wide range of $\Delta m_{15}$ values\footnote{The
value of $\Delta m_{15}$ refers to the difference between the $B$-band
SN brightness in magnitudes at maximum brightness and at
$+$15~rest-frame days.}  (see Phillips, M.~M. 1993) and the spectral
library provides a typical SN~Ia spectrum for each phase of the SN
lightcurve from $-10$ to $+70$ rest-frame days (Nugent~et~al. 2002).
We use these libraries to construct observed $VR$-band lightcurves
with a specified redshift and $\Delta m_{15}$ ranging from
0.8--1.9{\it mag} as follows.  By applying the $\Delta m_{15}$
weighting method of Prieto~et~al. (2006), we first construct $BVRI$
lightcurves for the specified $\Delta m_{15}$ value.  We then ``warp''
the spectrum to match the expected, rest-frame $B-V$ color at each
phase.  Finally we convolve the transmission curve of the $VR$ filter
(see Figure~\ref{fig:vrresp}) with the redshifted spectrum and obtain
the observed $VR$-band flux for a given phase.  We use a similar
procedure to construct reddened lightcurves.  After warping the
library spectrum to match the expected color for the specified $\Delta
m_{15}$ value, we approximate the host galaxy reddening by applying
the Cardelli~et~al.~(1989) Galactic reddening law using $R_{v}$ = 3.1
to the spectrum (see Riess et al. 1986b for discussion of host galaxy
reddening laws).  We then redshift the spectrum and apply the LMC
reddening law of Fitzpatrick (1986) with $R_{v}$ = 3.3.  We also add
the Galactic reddening using the Cardelli~et~al. law with $R_{v}$ =
3.1.  Finally, as in the unreddened case, we convolve the reddened,
redshifted spectrum with the $VR$-band transmission filter to obtain
the observed $VR$-band flux.  As with our own data, we normalize these
lightcurves to the flux at the time of maximum to create model $f_{\frac{VR}{VR_{max}}}$
lightcurves.

We use this lightcurve simulation tool in the following sections to
help us understand the impact of systematic effects on our findings.

\subsubsection{K-corrections} \label{section:kcorr}
The general K-correction formula (Schmidt et al. 1998, and Nugent et
al. 2002) is used to ``correct'' for the fact that, in a given filter,
observations of SNe with different redshifts sample different portions
of the SN~Ia rest-frame spectrum.  The observations are typically
normalized to the filter most closely matching the portion of the
rest-frame spectrum sampled by the filter in the observed frame.  To
apply such a correction to a given observation ideally requires a
spectrum taken at the same phase as the observation.  Because there
are few high-quality SN~Ia spectra prior to $-10$ rest-frame days, we
choose not to K-correct our lightcurves.  Instead, we choose SNe from
a narrow range of redshifts to avoid introducing scatter into our
composite by sampling very different portions of the SN~Ia spectrum.

To estimate the variation between the SNe in our bin, we construct
unreddened observed $VR$-band lightcurves at the redshifts of the SNe
in our composite using the lightcurve simulation tool described above.
We choose a fiducial $\Delta m_{15}$ of 1.2{\it mag} for these model
lightcurves.  Between $-10$ and $+80$ rest-frame days, the
flux/maximum flux ratio of the three lightcurves differs by less than
3\% with the maximum spread between the three at approximately $+$15
days.  All three lightcurves reach maximum brightness at the same
phase relative to rest-frame $B$-band maximum.  On the rising portion,
their $f_{\frac{VR}{VR_{max}}}$ lightcurves differ by less than 0.2\%.
These tests indicate that the systematic error contributed to an
estimate of the time to maximum brightness using a composite
lightcurve of SNe at redshifts between 0.135--0.165 without
K-corrections is negligible.

In addition to minimizing scatter between SNe at different redshifts,
K-corrections would provide a means for matching our observed
$VR$-band lightcurves to standard bands in the rest-frame.  At z=0.15,
the central redshift of the SNe in our composite lightcurve, the
observed $VR$-band most closely matches the rest-frame $V$-band.  To
compare the lightcurves of the observed $VR$-band at z=0.15 and
$V$-band at z=0, we construct $f_{\frac{VR}{VR_{max}}}$ lightcurves
between $-10$ and $+80$ rest-frame days with $\Delta m_{15}$ of
1.2{\it mag}.  Prior to maximum, the two lightcurves differ by
$\sim$2\% and reach maximum brightness at approximately the same phase
relative to $B$-band maximum.  Their times of maximum differ by less
than the resolution of our model lightcurves which is $\sim$0.5
rest-frame days.  Using a cubic spline fit to the lightcurves near
maximum, we find the difference in the times of maximum to be
1.1~rest-frame days.  Post-maximum, the lightcurves diverge with the
observed $VR$-band lightcurve declining more rapidly.  From this
comparison we conclude that for the rising portion of the lightcurve,
the observed $VR$-band lightcurve--with the time axis shifted to the
rest-frame--is a close approximation of the rest-frame $V$-band
lightcurve.  The systematic error in an estimate of the time to
$V$-band maximum using the observed $VR$-band lightcurve will be less
than $\pm$1.1~rest-frame days.

\subsubsection{SN~Ia Diversity}

Intrinsic diversity in the SN~Ia family will also impact our
estimate of the time-to-maximum from our composite lightcurve.  To
reduce the most gross impact of this effect, we remove the obvious
fast riser and decliner SM-2004-LMC-1060 from our composite
lightcurve.

\begin{figure}
\includegraphics[angle=0,width=8.4cm]{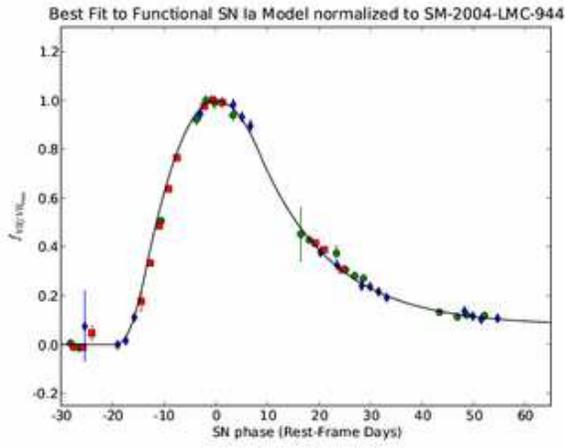}
\caption[Best Fit of Functional SN Ia Model to Composite Lightcurve]{The best fit of the functional SN~Ia model described in Section~\ref{section:templateLC} to the composite lightcurve.  Stretch parameters were added to normalize the width of the lightcurve to that of SM-2004-LMC-944 (squares).  The lightcurve shown has a $\Delta m_{-10}$ in the $VR$-band of 0.52{\it mag}.  For SM-2004-LMC-803 (circles), the stretch parameter is 1.01.  For SM-2004-LMC-797, the stretch parameter is 0.92.  Table~\ref{tab:parameters} gives the parameters and their uncertainties for this fit.}
\label{fig:bestfit}
\end{figure}

To account for the effect of variation between the remaining SNe, we
add a free ``stretch'' parameter, $s$, for each of the SNe in the fit
following Goldhaber et al. (2001).  Using C-MINUIT we perform a
multi-parameter fit to the composite lightcurve and fix the stretch
parameter for one of the SNe in the composite to 1, no stretch.
Effectively, the other SNe in the composite are normalized to the
shape of the unstretched SN.  We choose SM-2004-LMC-944 as our
fiducial SNe, because it has the median width of the 3 SNe in the
composite.  We present the results of this fit in
Table~\ref{tab:parameters}.  We characterize the ``shape'' of our best
fit by the value of $\Delta m_{-10}$, the difference in magnitudes
between the $VR$-band flux at $-10$ rest-frames days and at maximum.
For the best fit normalized to the shape of SM-2004-LMC-944, $\Delta
m_{-10}$ is 0.52{\it mag} and the time-to-maximum is 19.2$\pm$1.3~rest-frame
days.  Figure~\ref{fig:bestfit} shows the best fit with the composite
lightcurve.  The phases of the data points have been stretched
according to the values of $s$ returned by the best fit.
By scaling the time-to-maximum by the best-fit stretch parameters for
each of the other SNe in the composite, we can determine the
time-to-maximum for different values of $\Delta m_{-10}$. For
SM-2004-LMC-803 which has a $\Delta m_{-10}$ of 0.53{\it mag}, the
time-to-maximum is 18.96~rest-frame days.  For SM-2004-LMC-797 with a
$\Delta m_{-10}$ of 0.39{\it mag}, the time-to-maximum is
20.93~rest-frame days.

Our fits indicate that, like the declining portion of the lightcurve,
the shape of the rising lightcurve of a SN~Ia differs between
individual SNe in a way that can be paramaterized by a stretch factor.
With our current data, however, we cannot compare these rising
lightcurve shape parameters with those describing the decline rate.
This is because our lightcurves are not reddening corrected and, as
discussed below, the declining portion of our lightcurves is the most
sensitive to the impact of reddening.  Without reddening corrections
we cannot meaningfully compare the rate of rise with the rate of
decline in our lightcurves.  Further, because our $VR$-band lightcurve
differs most significantly from the standard, $V$-band filter on the
decline, comparing our findings to previous work in standard passbands
is also difficult.

\subsubsection{Reddening}
 
Reddening from dust along the line-of-sight to the SNe also alters the
shape of our composite lightcurve and impacts our estimates of the
parameters in our functional SN~Ia model, including the
time-to-maximum.  Because the SN spectrum evolves, the effect of
reddening changes with SN phase.  The bluer the intrinsic SN light,
the larger the change in the observed color caused by dust along the
line-of-sight.  The light from the SNe in our sample is reddened by
dust in three different locations: the host galaxy, the LMC, and the
Galaxy.  The line-of-sight dust introduces two different effects into
our composite lightcurve: 1) the overall change in the shape of the
composite lightcurve due to reddening and 2) increased scatter in the
composite lightcurve due to differences in the line-of-sight reddening
to the three separate SNe in the composite.

To examine the overall impact of reddening, we use the lightcurve
simulation tool described above to create an unreddened
$f_{\frac{VR}{VR_{max}}}$ lightcurve with $\Delta m_{15} =$ 1.2{\it
mag} at a redshift of 0.15.  We then create reddened lightcurves.  For
the host galaxy reddening we refer to the distribution of color
excesses found by the ESSENCE survey (Wood-Vasey, private
communication).  Assuming $R_{v}$ = 3.1, ESSENCE finds a mean value
for $E(B-V)$ of 0.06.  To obtain a reasonable estimate of $E(B-V)$
through the LMC, we double the mean value of the Galaxy-corrected
$E(B-V)$ for LMC stars found by Harris~et~al. (1997) and use
$E(B-V)$~=~0.26$\pm$0.055.  We also use $E(B-V)$~=~0.07$\pm$0.01
through the Galaxy toward the LMC as suggested by Harris~et~al. who
use the Oestreicher~et~al. (1995) SN1987A foreground reddening value.
We find the ratio of the reddened model lightcurve flux to the
unreddened model lightcurve flux at each phase, and multiply this
ratio by the data point in our composite lightcurve at the
corresponding phase.  For data points prior to $-10$ days, we use the
ratio at $-10$ days.  In this way we effectively ``redden'' our
composite lightcurve.  We find that in our $VR$-band at z=0.15, the
impact of reddening is significantly more severe on the declining arm
of the lightcurve.  The maximum change in $f_{\frac{VR}{VR_{max}}}$
due to reddening on the rising arm is $\sim$0.2\%, while the maximum
change on the declining arm is $\sim$5\%.  To get a more extreme
estimate of the impact of reddening, we also create a reddened
lightcurve with a host galaxy $E(B-V)$ of 0.25.  This value represents
approximately the 90th percentile host galaxy color excess found by
the ESSENCE survey.  Increasing the host galaxy reddening to this
amount can change the $f_{\frac{VR}{VR_{max}}}$ lightcurve by up to
$\sim$0.5\% on the rising arm and $\sim$10\% on the declining arm.
Because the rising arm of the lightcurve is so much less susceptible
to changes caused by reddening, we focus our analysis on the rising
portion of our composite lightcurve and the constraints we can place
on the time-to-maximum.

To understand how reddening impacts the value in the best fit of the
parameter $t_{r}$, we refit the reddened composite to our functional
model choosing SM-2004-LMC-944 as the fiducial SN for normalizing the
stretch.  We find that the estimate of the time-to-maximum, $-t_{r}$,
is increased by 1.6~rest-frame days.  From this we conclude that the
systematic shift in the time-to-maximum caused by reddening is
approximately $+$1.6~rest-frame days.  To examine whether the overall
effect of reddening is always to increase the time-to-maximum, we
choose extreme values for the color excess in the LMC,
$E(B-V)$~=~1.26, and the Galaxy, $E(B-V)$~=~1.07 and refit the
``reddened'' composite lightcurve.  As expected, the estimate of the
time-to-maximum is more significantly altered; the absolute value of
$t_{r}$ increases by almost 2~rest-frame days.  Notably, however, the
reddening only increases, and never decreases, the estimate of the
time-to-maximum.  From this we conclude that the overall potential
impact of reddening is to increase our estimate of the time-to-maximum
by 1.6~rest-frame days assuming reasonable values of the color excess
due to reddening.  We modify the value of our time-to-maximum to
reflect the impact of the reddening to obtain a best estimate of
17.6$\pm$1.3~rest-frame days.

We use Monte Carlo simulations to examine how the uncertainties in
the LMC, Galaxy, and host galaxy reddenings as well as the differences
between host galaxy reddening for each of the SNe impact our estimate
of the overall effect of reddening.  For each simulation, we create
multiple realizations of a reddened composite lightcurve in the manner
described above.  We perform a multi-parameter fit on each realization
and calculate the robust mean value of the time-to-maximum and its
standard deviation.  To isolate the effect of the uncertainty in each
source of reddening, we hold the color excess values of the other
reddening sources fixed and vary the source of interest.  For example,
to understand how the uncertainty in the LMC's color excess affects
our estimate of the impact of reddening, we set the Galaxy's $E(B-V)$
to 0.07 and the host galaxy $E(B-V)$ for all three SNe to 0.06.  For
each realization, we draw the LMC's color excess from a gaussian
distribution with a mean of 0.26 and a $\sigma$ of 0.55, reflecting
the values determined by Harris~et~al.  We perform a similar Monte
Carlo holding the LMC and host galaxy reddenings fixed while choosing
the Galactic color excess from a gaussian distribution centered at
0.07 with a $\sigma$ of 0.01.  Finally, we estimate the combined
impact of our uncertainty in the host galaxy reddening values and the
differences between them for each SNe.  Holding the LMC and Galactic
reddening fixed in each realization, we choose a different host galaxy
color excess for each SN from a distribution of host galaxy $E(B-V)$
similar to that found by the ESSENCE survey.  

For each of the simulations described above, the 3$\sigma$-clipped
mean value of the time-to-maximum matched that obtained by using the
``best guess'' values of the reddenings.  The standard deviations
about this mean provides an estimate of the systematic uncertainty in
our reddening-corrected time-to-maximum caused by uncertainties in the
reddening caused by each source.  For the LMC, the standard deviation
of the time-to-maximum is 0.014.  For the Galaxy the standard
deviation is 0.012.  For the host galaxy reddenings the standard
deviation is 0.067.  Summing these numbers in quadrature, we arrive at
an estimate of the total systematic uncertainty in the time-to-maximum
due to reddening, $\pm$0.07~rest-frame days.

\subsection{Comparison with Previous Findings} 

Our investigation of systematic effects impacting our composite
lightcurve yields the following conclusions.  The lack of
K-corrections on our SNe chosen from the narrow redshift range of
0.135--0.165 will have a negligible effect on the overall shape of our
composite lightcurve.  Without K-corrections, however, we must be
careful in how we compare our observed $VR$-band lightcurve with the
most closely matched rest-frame filter, the $V$-band.  We find that
the rising portion of our observed $VR$-band lightcurve is similar to
the rest-frame $V$-band, and that an estimate of the time-of-maximum
from our composite lightcurve will differ from the $V$-band
time-of-maximum by less than 1.1~rest-frame days.  To account for
intrinsic variability we introduce a stretch parameter for each of the
SNe in the composite lightcurve and normalize the shape to
SM-2004-LMC-944.  We estimate that the overall effect of reddening on
the time-to-maximum is to increase it by 1.6~rest-frame days.  The
systematic error in our estimate of the effect of reddening is
$\pm$0.07~rest-frame days.

Based on the fits described above, the best fit parameters to our
functional model give a time-of-explosion 17.6$\pm$1.3({\it
stat})$\pm$0.07({\it sys})~rest-frame days before maximum $VR$-band
brightness for a SN~Ia with a $\Delta m_{-10}$ of 0.52{\it mag}.  At a
z of 0.15, we expect the observed $VR$-band to most closely match the
rest-frame $V$-band lightcurve, and we add an additional systematic
uncertainty of $\pm$1.1~rest-frame days to our estimate of the
time-to-maximum in the $V$-band.  Our findings give a smaller value
for the time-to-maximum than that of Riess et al. (1999) for the
fiducial $V$-band who find a time-to-maximum of 21.1$\pm$0.2 days.
The significance of this discrepancy is unclear.  Our value for the
time-to-maximum is normalized to an SN with $VR$-band $\Delta
m_{-10}$~=~0.52{\it mag}.  As described above, comparing our values of
$\Delta m_{-10}$ with previous work is difficult.  For this paper we
note the discrepency but without a study that analyzes both our
lightcurves and previous data in the same way, we cannot comment on
its significance.

\section{Conclusion}
We present $VR$-band lightcurves and optical spectra of 11 SNe~Ia
behind the LMC discovered by the SuperMACHO survey\footnote{See
http://ctiokw.ctio.noao.edu/$\sim$sm/sm/SNrise for electronic data
tables}.  Our data include some of the earliest pre-maximum detections
of SNe~Ia.  We provide a functional model for the observed $VR$-band
lightcurve from the time of explosion to $+60$ days by fitting a
composite lightcurve to three SNe in the redshift bin of
z=0.135--0.165.  The data are fitted without K-corrections or
reddening corrections; however, the set of SNe have been chosen to
minimize the impact of these effects.  Our function uses the expanding
fireball model of Goldhaber et al. (1998) to describe the lightcurve
immediately following the explosion.  The best fit of our functional
model to our composite, observed $VR$-band lightcurve gives a
time-to-maximum of 17.6$\pm$1.3({\it stat})$\pm$0.07({\it
sys})~rest-frame days for a SN~Ia with a $\Delta m_{-10}$ of 0.52{\it
mag}.  Our simulations indicate that the $VR$-band time-of-maximum at
z=0.15 should match the rest-frame $V$-band time-of-maximum to within
1.1~rest-frame days.

We present these data to be used to test competing models of the SN~Ia
explosion mechanism by placing observational limits on the time to
maximum and the shape of the rising lightcurve.  Analyses of our data
are limited by its being in a single band.  While our broadband filter
enables us to detect flux earlier, we cannot calibrate our lightcurves
against the nearby sets of SNe~Ia observed in $BVRI$.\footnote{Because
the transmission curve of our $VR$ filter differs from the sum of the
$V$-band and $R$-band transmissions, we cannot simply add the nearby
$V$ and $R$ templates to obtain a $VR$ template.}  An ideal study
should include both a broadband filter and the standard filter set.

\section{Acknowledgments}

The SuperMACHO survey is being undertaken under the auspices of the
NOAO Survey Program. We are very grateful for the support provided to
the Survey program from the NOAO and the National Science
Foundation. We are particularly indebted to the scientists and staff
at the Cerro Tololo Interamerican Observatory for their assistance in
helping us carry out the SuperMACHO survey. We also appreciate the
invaluable help of Mr.  Chance Reschke in building and maintaining the
computing cluster we use for image analysis. SuperMACHO is supported
by the STScI grant GO-10583.  This project works closely with members
of the ESSENCE supernova survey, and we are grateful for their input
and assistance. The spectroscopic observations presented in this paper
were obtained on the Magellan telescopes operated by the Las Campanas
Observatory.  We are grateful to the scientists and staff of
LCO. Discussions with P.~Pinto were invaluable in helping us
understand the physics underlying our observations.  We are also
grateful to T.~Matheson and M.~Modjaz for their help with spectral
identifications.  The support of the McDonnell Foundation, through a
Centennial Fellowship awarded to C. Stubbs, has been essential to the
SuperMACHO survey. We are most grateful for the Foundation's support
for this project.  C.~Stubbs and AG are also grateful for support from
Harvard University.  AG would like to thank the University of
Washington Depratment of Astronomy for facilities support.  KHC's,
MEH's, and SN's work was performed under the auspices of the U.S.
Department of Energy, National Nuclear Security Administration by the
University of California, Lawrence Livermore National Laboratory under
contract No. W-7405-Eng-48.  JLP's work is supported by STScI grant
GO-9860.07.  AC acknowledges the support of CONICYT (Chile) through
FONDECYT grant 1051061.  DM and LM acknowledge support from the Fondap
Center for Astrophysics grant 15010003. DLW acknowledges financial
support in the form of a Discovery Grant from the Natural Sciences and
Engineering Research Council of Canada (NSERC).



\clearpage

\begin{deluxetable}{lcccccrrr}
\tabletypesize{\scriptsize}
\tablecaption{SuperMACHO Supernovae 2004}
\tablewidth{0pt}
\rotate
\tablehead{
\colhead{SN ID} & \colhead{RA (J2000)} & \colhead{DEC} & \colhead{z} & \colhead{galaxy z} & \colhead{phase$_{S/N>5}$} & \colhead{$t_{max} (MJD)$} & \colhead{$VR_{max}$} & \colhead{phase$_{f=0}$}
}
\startdata
SM-2004-LMC-64\tablenotemark{a} & 04:55:22.266 & -67:30:44.31 & 0.22 & \nodata & -7.9 & 53292.97$\pm$0.86 & 64.93$\pm$1.04 & -29.3\\
SM-2004-LMC-772 & 05:19:42.656 & -67:31:35.83 & 0.19 & \nodata & -18.0 & 53316.74$\pm$0.39 & 79.83$\pm$1.13 & -566.8\\
SM-2004-LMC-797 & 05:59:13.224 & -71:49:59.27 & 0.145 & \nodata & -17.2 & 53318.94$\pm$1.00 & 96.05$\pm$1.42 &-20.7\\
SM-2004-LMC-803 & 05:47:05.071 & -71:46:28.36 & 0.16 & \nodata & -10.4 & 53327.46$\pm$0.53 & 69.86$\pm$0.87 & -27.8\\
SM-2004-LMC-811 & 04:56:31.608 & -66:58:09.21 & 0.27 & \nodata & -7.6 & 53324.87$\pm$0.97 & 31.12$\pm$0.62 & -20.2\\
SM-2004-LMC-917 & 05:21:19.819 & -70:51:12.57 & 0.11 & \nodata & -5.5 & 53350.52$\pm$0.28 & 198.76$\pm$0.56 & -24.6\\
SM-2004-LMC-944 & 05:11:48.947 & -70:29:38.66 & 0.15 & \nodata & -12.7 & 53358.87$\pm$0.50 & 60.49$\pm$0.49 & -37.9\\
SM-2004-LMC-1002 & 04:53:09.337 & -69:41:00.13 & 0.35 & 0.350 & -8.8 & 53356.12$\pm$15.44 & 14.93$\pm$3.24 & -30.3\\
SM-2004-LMC-1052 & 06:01:36.188 & -71:59:29.88 & 0.34 & 0.348 & -9.5 & 53361.10$\pm$2.81 & 17.09$\pm$0.84 & -22.2\\
SM-2004-LMC-1060 & 05:35:30.148 & -71:06:34.05 & 0.16 & 0.154 & -13.5 & 53363.94$\pm$1.96 & 76.73$\pm$3.60 & -326.4\\
SM-2004-LMC-1102 & 05:37:13.676 & -68:50:00.93 & 0.22 & \nodata & -13.1 & 53364.30$\pm$1.22 & 31.65$\pm$1.17 & -27.0\\
\enddata
\tablenotetext{a}{SM-LMC-2004-64 also has IAU designation SN2004gb.}
\tablecomments{Summary of SNe~Ia presented in this paper.  {\it SN~ID} gives the SuperMACHO survey identification of each SN.  {\it z} is the redshift of the SN determined by comparing its spectrum to a nearby SN.  {\it Galaxy~z} is the redshift of the SN's host galaxy determined, when possible, from galaxy features in the spectrum.  {\it phase$_{S/N>5}$} indicates the rest-frame phase in days at which the first detection with S/N $>$ 5 was made.  $t_{max}$ is the time of $VR$-band maximum, $VR_{max}$.  Both $t_{max}$ and $VR_{max}$ are given with their 1$\sigma$ uncertainties.  $VR_{max}$ is in flux units normalized to a zeropoint of 25.  {\it phase$_{f=0}$} gives the rest-frame phase of the last zero flux measurement, corresponding to difference flux with S/N$<$0.5, prior to the SN's detection.}
\label{tab:SNinfo}
\end{deluxetable}

\clearpage
\begin{deluxetable}{ccccccc}
\tabletypesize{\scriptsize}
\tablecaption{Lightcurve for SM-2004-LMC-64}
\tablewidth{0pt}
\tablehead{
\colhead{MJD} & \colhead{Rest Phase} & \colhead{Diff flux} & \colhead{flux err} & \colhead{$f_{\frac{VR}{VR_{max}}}$} & \colhead{$f_{\frac{VR}{VR_{max}}}$ err} & \colhead{S/N}
}
\startdata
53257.27 & -29.3 & -0.410 & 1.168 & -0.006 & 2.852 & 0.35 \\
53266.23 & -21.9 & -3.083 & 2.099 & -0.047 & 0.681 & 1.47 \\
53283.31 & -7.9 & 40.694 & 2.666 & 0.627 & 0.067 & 15.26 \\
53287.37 & -4.6 & 54.304 & 5.662 & 0.836 & 0.105 & 9.59 \\
53289.30 & -3.0 & 60.166 & 1.196 & 0.927 & 0.026 & 50.31 \\
53291.20 & -1.4 & 66.107 & 1.746 & 1.018 & 0.031 & 37.86 \\
53293.18 &  0.2 & 68.740 & 2.673 & 1.059 & 0.042 & 25.71 \\
53297.30 &  3.5 & 60.655 & 1.158 & 0.934 & 0.025 & 52.37 \\
53299.19 &  5.1 & 58.575 & 1.269 & 0.902 & 0.027 & 46.15 \\
53301.18 &  6.7 & 53.818 & 1.580 & 0.829 & 0.033 & 34.06 \\
53321.21 & 23.1 & 20.107 & 1.633 & 0.310 & 0.083 & 12.31 \\
53325.29 & 26.5 & 16.623 & 1.055 & 0.256 & 0.065 & 15.76 \\
53327.33 & 28.2 & 16.610 & 0.894 & 0.256 & 0.056 & 18.58 \\
53331.32 & 31.4 & 12.894 & 0.930 & 0.199 & 0.074 & 13.87 \\
53344.29 & 42.1 & 7.851 & 0.834 & 0.121 & 0.107 & 9.41 \\
53348.36 & 45.4 & 7.481 & 2.565 & 0.115 & 0.343 & 2.92 \\
53350.20 & 46.9 & 6.244 & 0.692 & 0.096 & 0.112 & 9.02 \\
53352.31 & 48.6 & 9.055 & 2.524 & 0.139 & 0.279 & 3.59 \\
53354.26 & 50.2 & 6.311 & 0.891 & 0.097 & 0.142 & 7.08 \\
53356.26 & 51.9 & 6.288 & 0.849 & 0.097 & 0.136 & 7.41 \\
\enddata
\tablecomments{Difference flux lightcurve for SM-2004-LMC-64. Rest Phase is given in rest-frame days relative to observed $VR$-band maximum.  {\it Diff~flux} is the observed $VR$-band difference flux at the position of the SN given in Table~\ref{tab:SNinfo}.  These fluxes are determined using the N(N-1)/2 method of Barris et al. (2005) and are normalized to a zeropoint of 25.  {\it Flux~err} is the error in {\it Diff~flux}.  $f_{\frac{VR}{VR_{max}}}$ is the difference flux normalized by the maxixmum $VR$-band flux, {\it $VR_{max}$}, given in Table~\ref{tab:SNinfo}.  $f_{\frac{VR}{VR_{max}}}$~err is the error in $f_{\frac{VR}{VR_{max}}}$ and includes the uncertainty in $VR_{max}$.  {\it S/N} gives the significance of the difference flux measurement.}
\label{tab:2004-LMC-64.tab}
\end{deluxetable}

\begin{deluxetable}{ccccccc}
\tabletypesize{\scriptsize}
\tablecaption{Lightcurve for SM-2004-LMC-772}
\tablewidth{0pt}
\tablehead{
\colhead{MJD} & \colhead{Rest Phase} & \colhead{Diff flux} & \colhead{flux err} & \colhead{$f_{\frac{VR}{VR_{max}}}$} & \colhead{$f_{\frac{VR}{VR_{max}}}$ err} & \colhead{S/N}
}
\startdata
53257.37 & -49.9 & 1.176 & 0.995 & 0.015 & 0.846 & 1.18 \\
53289.35 & -23.0 & 1.866 & 1.074 & 0.023 & 0.576 & 1.74 \\
53295.28 & -18.0 & 7.319 & 0.994 & 0.092 & 0.137 & 7.36 \\
53315.30 & -1.2 & 78.786 & 1.445 & 0.987 & 0.023 & 54.52 \\
53323.27 &  5.5 & 64.686 & 1.135 & 0.810 & 0.023 & 56.99 \\
53327.35 &  8.9 & 52.777 & 1.238 & 0.661 & 0.027 & 42.61 \\
53329.36 & 10.6 & 48.186 & 1.250 & 0.604 & 0.030 & 38.56 \\
53344.35 & 23.2 & 25.464 & 1.204 & 0.319 & 0.049 & 21.16 \\
53346.34 & 24.9 & 22.006 & 1.346 & 0.276 & 0.063 & 16.35 \\
53348.24 & 26.5 & 21.588 & 0.850 & 0.270 & 0.042 & 25.40 \\
53350.29 & 28.2 & 20.707 & 1.089 & 0.259 & 0.054 & 19.02 \\
53352.22 & 29.8 & 18.823 & 1.145 & 0.236 & 0.062 & 16.44 \\
53354.22 & 31.5 & 17.099 & 1.136 & 0.214 & 0.068 & 15.05 \\
53356.24 & 33.2 & 16.935 & 0.655 & 0.212 & 0.041 & 25.86 \\
53358.32 & 34.9 & 14.938 & 0.910 & 0.187 & 0.063 & 16.41 \\
53360.28 & 36.6 & 13.548 & 0.710 & 0.170 & 0.054 & 19.08 \\
53379.13 & 52.4 & 8.437 & 0.733 & 0.106 & 0.088 & 11.51 \\
53381.15 & 54.1 & 7.082 & 0.762 & 0.089 & 0.109 & 9.29 \\
53383.15 & 55.8 & 7.653 & 0.600 & 0.096 & 0.080 & 12.75 \\
53387.13 & 59.2 & 7.282 & 1.002 & 0.091 & 0.138 & 7.27 \\
\enddata
\tablecomments{Difference flux lightcurve for SM-2004-LMC-772. See Table~\ref{tab:2004-LMC-64.tab} for explanation of column headings.}
\label{tab:2004-LMC-772.tab}
\end{deluxetable}

\begin{deluxetable}{ccccccc}
\tabletypesize{\scriptsize}
\tablecaption{Lightcurve for SM-2004-LMC-797}
\tablewidth{0pt}
\tablehead{
\colhead{MJD} & \colhead{Rest Phase} & \colhead{Diff flux} & \colhead{flux err} & \colhead{$f_{\frac{VR}{VR_{max}}}$} & \colhead{$f_{\frac{VR}{VR_{max}}}$ err} & \colhead{S/N}
}
\startdata
53287.28 & -27.7 & 7.218 & 13.565 & 0.075 & 1.879 & 0.53 \\
53295.24 & -20.7 & -0.123 & 0.930 & -0.001 & 7.533 & 0.13 \\
53297.19 & -19.0 & 1.495 & 0.946 & 0.016 & 0.633 & 1.58 \\
53299.27 & -17.2 & 10.651 & 0.770 & 0.111 & 0.074 & 13.82 \\
53315.20 & -3.3 & 90.763 & 1.869 & 0.945 & 0.025 & 48.56 \\
53323.21 &  3.7 & 94.325 & 1.822 & 0.982 & 0.024 & 51.78 \\
53325.30 &  5.6 & 89.474 & 1.775 & 0.932 & 0.025 & 50.41 \\
53327.36 &  7.4 & 85.997 & 2.309 & 0.895 & 0.031 & 37.24 \\
53344.28 & 22.1 & 36.229 & 1.096 & 0.377 & 0.034 & 33.06 \\
53348.31 & 25.7 & 31.168 & 0.843 & 0.324 & 0.031 & 36.99 \\
53354.26 & 30.9 & 23.170 & 1.838 & 0.241 & 0.081 & 12.60 \\
53356.28 & 32.6 & 22.785 & 0.817 & 0.237 & 0.039 & 27.90 \\
53358.34 & 34.4 & 20.772 & 1.042 & 0.216 & 0.052 & 19.94 \\
53360.30 & 36.1 & 18.483 & 0.768 & 0.192 & 0.044 & 24.06 \\
53379.13 & 52.6 & 13.162 & 0.969 & 0.137 & 0.075 & 13.58 \\
53381.15 & 54.3 & 11.145 & 0.942 & 0.116 & 0.086 & 11.83 \\
53383.15 & 56.1 & 9.917 & 0.684 & 0.103 & 0.071 & 14.50 \\
53387.12 & 59.5 & 10.317 & 0.966 & 0.107 & 0.095 & 10.69 \\
\enddata
\tablecomments{Difference flux lightcurve for SM-2004-LMC-797. See Table~\ref{tab:2004-LMC-64.tab} for explanation of column headings.}
\label{tab:2004-LMC-797.tab}
\end{deluxetable}

\begin{deluxetable}{ccccccc}
\tabletypesize{\scriptsize}
\tablecaption{Lightcurve for SM-2004-LMC-803}
\tablewidth{0pt}
\tablehead{
\colhead{MJD} & \colhead{Rest Phase} & \colhead{Diff flux} & \colhead{flux err} & \colhead{$f_{\frac{VR}{VR_{max}}}$} & \colhead{$f_{\frac{VR}{VR_{max}}}$ err} & \colhead{S/N}
}
\startdata
53295.23 & -27.8 & 0.361 & 1.092 & 0.005 & 3.029 & 0.33 \\
53297.18 & -26.1 & -0.886 & 1.037 & -0.013 & 1.170 & 0.85 \\
53315.34 & -10.4 & 35.393 & 1.081 & 0.507 & 0.033 & 32.74 \\
53323.21 & -3.7 & 64.359 & 1.536 & 0.921 & 0.027 & 41.91 \\
53325.30 & -1.9 & 69.728 & 1.529 & 0.998 & 0.025 & 45.60 \\
53327.23 & -0.2 & 69.143 & 1.636 & 0.990 & 0.027 & 42.26 \\
53331.30 &  3.3 & 65.585 & 1.389 & 0.939 & 0.025 & 47.21 \\
53346.37 & 16.3 & 31.534 & 7.572 & 0.451 & 0.240 & 4.16 \\
53348.27 & 17.9 & 29.872 & 0.814 & 0.428 & 0.030 & 36.72 \\
53354.27 & 23.1 & 26.101 & 2.085 & 0.374 & 0.081 & 12.52 \\
53356.28 & 24.8 & 21.420 & 0.823 & 0.307 & 0.040 & 26.02 \\
53358.33 & 26.6 & 19.618 & 1.053 & 0.281 & 0.055 & 18.62 \\
53360.25 & 28.3 & 18.968 & 0.820 & 0.272 & 0.045 & 23.12 \\
53377.14 & 42.8 & 9.255 & 0.696 & 0.132 & 0.076 & 13.29 \\
53381.12 & 46.3 & 7.939 & 0.971 & 0.114 & 0.123 & 8.18 \\
53383.12 & 48.0 & 8.478 & 0.661 & 0.121 & 0.079 & 12.82 \\
53387.15 & 51.5 & 8.239 & 0.985 & 0.118 & 0.120 & 8.37 \\
\enddata
\tablecomments{Difference flux lightcurve for SM-2004-LMC-803. See Table~\ref{tab:2004-LMC-64.tab} for explanation of column headings.}
\label{tab:2004-LMC-803.tab}
\end{deluxetable}

\begin{deluxetable}{ccccccc}
\tabletypesize{\scriptsize}
\tablecaption{Lightcurve for SM-2004-LMC-811}
\tablewidth{0pt}
\tablehead{
\colhead{MJD} & \colhead{Rest Phase} & \colhead{Diff flux} & \colhead{flux err} & \colhead{$f_{\frac{VR}{VR_{max}}}$} & \colhead{$f_{\frac{VR}{VR_{max}}}$ err} & \colhead{S/N}
}
\startdata
53287.37 & -29.5 & 17.718 & 14.242 & 0.569 & 0.804 & 1.24 \\
53289.31 & -28.0 & -1.034 & 0.705 & -0.033 & 0.682 & 1.47 \\
53295.19 & -23.4 & 0.212 & 0.779 & 0.007 & 3.681 & 0.27 \\
53297.30 & -21.7 & -0.317 & 0.795 & -0.010 & 2.507 & 0.40 \\
53299.19 & -20.2 & 0.174 & 0.830 & 0.006 & 4.763 & 0.21 \\
53301.18 & -18.7 & -1.227 & 1.016 & -0.039 & 0.828 & 1.21 \\
53315.25 & -7.6 & 23.727 & 1.111 & 0.762 & 0.051 & 21.36 \\
53321.21 & -2.9 & 31.035 & 3.161 & 0.997 & 0.104 & 9.82 \\
53325.29 &  0.3 & 31.162 & 1.091 & 1.001 & 0.040 & 28.55 \\
53327.33 &  1.9 & 30.392 & 1.117 & 0.977 & 0.042 & 27.21 \\
53331.32 &  5.1 & 30.126 & 1.164 & 0.968 & 0.044 & 25.88 \\
53344.29 & 15.3 & 18.166 & 1.125 & 0.584 & 0.065 & 16.15 \\
53348.36 & 18.5 & 7.853 & 2.961 & 0.252 & 0.378 & 2.65 \\
53350.20 & 19.9 & 12.294 & 0.695 & 0.395 & 0.060 & 17.68 \\
53352.31 & 21.6 & 10.440 & 1.105 & 0.335 & 0.108 & 9.45 \\
53354.17 & 23.1 & 9.706 & 0.986 & 0.312 & 0.103 & 9.85 \\
53360.35 & 27.9 & 6.264 & 0.976 & 0.201 & 0.157 & 6.42 \\
53385.14 & 47.5 & 3.283 & 0.671 & 0.105 & 0.205 & 4.89 \\
53387.14 & 49.0 & 3.416 & 0.886 & 0.110 & 0.260 & 3.85 \\
\enddata
\tablecomments{Difference flux lightcurve for SM-2004-LMC-811. See Table~\ref{tab:2004-LMC-64.tab} for explanation of column headings.}
\label{tab:2004-LMC-811.tab}
\end{deluxetable}

\begin{deluxetable}{ccccccc}
\tabletypesize{\scriptsize}
\tablecaption{Lightcurve for SM-2004-LMC-917}
\tablewidth{0pt}
\tablehead{
\colhead{MJD} & \colhead{Rest Phase} & \colhead{Diff flux} & \colhead{flux err} & \colhead{$f_{\frac{VR}{VR_{max}}}$} & \colhead{$f_{\frac{VR}{VR_{max}}}$ err} & \colhead{S/N}
}
\startdata
53295.27 & -49.8 & 0.075 & 0.588 & 0.000 & 7.801 & 0.13 \\
53297.22 & -48.0 & 1.027 & 0.673 & 0.005 & 0.655 & 1.53 \\
53301.24 & -44.4 & 0.584 & 1.077 & 0.003 & 1.844 & 0.54 \\
53315.22 & -31.8 & -0.731 & 0.660 & -0.004 & 0.903 & 1.11 \\
53323.20 & -24.6 & 0.248 & 0.503 & 0.001 & 2.027 & 0.49 \\
53325.36 & -22.7 & 0.880 & 1.209 & 0.004 & 1.374 & 0.73 \\
53329.29 & -19.1 & -0.947 & 0.626 & -0.005 & 0.661 & 1.51 \\
53331.33 & -17.3 & 3.916 & 0.790 & 0.020 & 0.202 & 4.96 \\
53344.36 & -5.6 & 163.411 & 2.276 & 0.822 & 0.014 & 71.79 \\
53346.35 & -3.8 & 181.941 & 2.037 & 0.915 & 0.012 & 89.34 \\
53348.19 & -2.1 & 192.288 & 1.344 & 0.967 & 0.008 & 143.09 \\
53350.27 & -0.2 & 199.294 & 0.684 & 1.003 & 0.004 & 291.32 \\
53352.33 &  1.6 & 195.035 & 1.130 & 0.981 & 0.006 & 172.60 \\
53360.35 &  8.9 & 153.213 & 2.068 & 0.771 & 0.014 & 74.08 \\
53385.17 & 31.2 & 53.737 & 1.428 & 0.270 & 0.027 & 37.64 \\
\enddata
\tablecomments{Difference flux lightcurve for SM-2004-LMC-917. See Table~\ref{tab:2004-LMC-64.tab} for explanation of column headings.}
\label{tab:2004-LMC-917.tab}
\end{deluxetable}

\begin{deluxetable}{ccccccc}
\tabletypesize{\scriptsize}
\tablecaption{Lightcurve for SM-2004-LMC-944}
\tablewidth{0pt}
\tablehead{
\colhead{MJD} & \colhead{Rest Phase} & \colhead{Diff flux} & \colhead{flux err} & \colhead{$f_{\frac{VR}{VR_{max}}}$} & \colhead{$f_{\frac{VR}{VR_{max}}}$ err} & \colhead{S/N}
}
\startdata
53315.26 & -37.9 & 0.070 & 0.778 & 0.001 & 11.177 & 0.09 \\
53323.34 & -30.9 & 0.609 & 0.898 & 0.010 & 1.475 & 0.68 \\
53327.22 & -27.5 & -0.557 & 0.851 & -0.009 & 1.528 & 0.65 \\
53329.27 & -25.7 & -0.563 & 0.689 & -0.009 & 1.222 & 0.82 \\
53331.23 & -24.0 & 2.834 & 1.696 & 0.047 & 0.598 & 1.67 \\
53342.24 & -14.5 & 10.673 & 2.345 & 0.176 & 0.220 & 4.55 \\
53344.24 & -12.7 & 20.172 & 0.552 & 0.333 & 0.029 & 36.53 \\
53346.32 & -10.9 & 29.423 & 0.661 & 0.486 & 0.024 & 44.53 \\
53348.35 & -9.1 & 38.566 & 0.635 & 0.638 & 0.018 & 60.72 \\
53350.24 & -7.5 & 46.252 & 0.739 & 0.765 & 0.018 & 62.60 \\
53356.33 & -2.2 & 59.033 & 0.557 & 0.976 & 0.012 & 105.97 \\
53358.18 & -0.6 & 60.552 & 0.768 & 1.001 & 0.015 & 78.88 \\
53360.32 &  1.3 & 59.972 & 1.167 & 0.991 & 0.021 & 51.38 \\
53381.12 & 19.3 & 25.134 & 0.623 & 0.416 & 0.026 & 40.36 \\
53383.12 & 21.1 & 23.555 & 0.521 & 0.389 & 0.024 & 45.17 \\
53387.15 & 24.6 & 18.505 & 0.621 & 0.306 & 0.035 & 29.81 \\
\enddata
\tablecomments{Difference flux lightcurve for SM-2004-LMC-944. See Table~\ref{tab:2004-LMC-64.tab} for explanation of column headings.}
\label{tab:2004-LMC-944.tab}
\end{deluxetable}

\begin{deluxetable}{ccccccc}
\tabletypesize{\scriptsize}
\tablecaption{Lightcurve for SM-2004-LMC-1002}
\tablewidth{0pt}
\tablehead{
\colhead{MJD} & \colhead{Rest Phase} & \colhead{Diff flux} & \colhead{flux err} & \colhead{$f_{\frac{VR}{VR_{max}}}$} & \colhead{$f_{\frac{VR}{VR_{max}}}$ err} & \colhead{S/N}
}
\startdata
53295.18 & -45.1 & -0.021 & 0.828 & -0.001 & 40.005 & 0.02 \\
53297.30 & -43.6 & -0.590 & 0.738 & -0.040 & 1.269 & 0.80 \\
53299.19 & -42.2 & -1.349 & 1.676 & -0.090 & 1.261 & 0.80 \\
53301.17 & -40.7 & -1.034 & 1.109 & -0.069 & 1.095 & 0.93 \\
53315.24 & -30.3 & 0.282 & 0.822 & 0.019 & 2.922 & 0.34 \\
53325.28 & -22.8 & 0.576 & 0.746 & 0.039 & 1.312 & 0.77 \\
53327.32 & -21.3 & -1.888 & 0.747 & -0.126 & 0.451 & 2.53 \\
53329.34 & -19.8 & -0.643 & 0.541 & -0.043 & 0.868 & 1.19 \\
53331.35 & -18.3 & -0.582 & 1.241 & -0.039 & 2.144 & 0.47 \\
53344.31 & -8.7 & 9.045 & 1.217 & 0.606 & 0.256 & 7.43 \\
53346.28 & -7.3 & 13.110 & 1.145 & 0.878 & 0.234 & 11.45 \\
53348.22 & -5.9 & 12.412 & 0.646 & 0.831 & 0.223 & 19.22 \\
53350.19 & -4.4 & 13.807 & 0.676 & 0.925 & 0.223 & 20.44 \\
53352.26 & -2.9 & 13.994 & 0.736 & 0.937 & 0.224 & 19.01 \\
53354.18 & -1.4 & 15.154 & 0.914 & 1.015 & 0.225 & 16.59 \\
53360.37 &  3.1 & 11.610 & 3.738 & 0.778 & 0.388 & 3.11 \\
53385.14 & 21.5 & 2.049 & 0.674 & 0.137 & 0.394 & 3.04 \\
\enddata
\tablecomments{Difference flux lightcurve for SM-2004-LMC-1002. See Table~\ref{tab:2004-LMC-64.tab} for explanation of column headings.}
\label{tab:2004-LMC-1002.tab}
\end{deluxetable}

\begin{deluxetable}{ccccccc}
\tabletypesize{\scriptsize}
\tablecaption{Lightcurve for SM-2004-LMC-1052}
\tablewidth{0pt}
\tablehead{
\colhead{MJD} & \colhead{Rest Phase} & \colhead{Diff flux} & \colhead{flux err} & \colhead{$f_{\frac{VR}{VR_{max}}}$} & \colhead{$f_{\frac{VR}{VR_{max}}}$ err} & \colhead{S/N}
}
\startdata
53295.24 & -49.2 & -0.666 & 0.843 & -0.039 & 1.265 & 0.79 \\
53297.19 & -47.7 & 0.583 & 0.952 & 0.034 & 1.634 & 0.61 \\
53299.27 & -46.1 & 0.859 & 0.752 & 0.050 & 0.876 & 1.14 \\
53315.20 & -34.3 & 1.078 & 1.180 & 0.063 & 1.096 & 0.91 \\
53323.21 & -28.3 & 2.211 & 0.847 & 0.129 & 0.386 & 2.61 \\
53325.30 & -26.7 & 0.144 & 0.838 & 0.008 & 5.810 & 0.17 \\
53327.36 & -25.2 & 0.578 & 1.357 & 0.034 & 2.348 & 0.43 \\
53331.30 & -22.2 & -0.349 & 1.339 & -0.020 & 3.836 & 0.26 \\
53344.28 & -12.6 & 3.058 & 0.867 & 0.179 & 0.288 & 3.53 \\
53348.31 & -9.5 & 8.714 & 0.673 & 0.510 & 0.092 & 12.95 \\
53352.34 & -6.5 & 10.739 & 1.842 & 0.628 & 0.179 & 5.83 \\
53354.26 & -5.1 & 16.207 & 1.952 & 0.948 & 0.130 & 8.30 \\
53356.28 & -3.6 & 16.779 & 0.984 & 0.982 & 0.077 & 17.05 \\
53358.34 & -2.1 & 16.838 & 0.954 & 0.985 & 0.075 & 17.65 \\
53360.30 & -0.6 & 16.303 & 0.907 & 0.954 & 0.074 & 17.98 \\
53379.13 & 13.5 & 7.213 & 0.908 & 0.422 & 0.135 & 7.95 \\
53381.15 & 15.0 & 5.784 & 1.104 & 0.338 & 0.197 & 5.24 \\
53383.15 & 16.5 & 4.348 & 0.750 & 0.254 & 0.179 & 5.80 \\
53387.12 & 19.4 & 2.766 & 1.148 & 0.162 & 0.418 & 2.41 \\
\enddata
\tablecomments{Difference flux lightcurve for SM-2004-LMC-1052. See Table~\ref{tab:2004-LMC-64.tab} for explanation of column headings.}
\label{tab:2004-LMC-1052.tab}
\end{deluxetable}

\begin{deluxetable}{ccccccc}
\tabletypesize{\scriptsize}
\tablecaption{Lightcurve for SM-2004-LMC-1060}
\tablewidth{0pt}
\tablehead{
\colhead{MJD} & \colhead{Rest Phase} & \colhead{Diff flux} & \colhead{flux err} & \colhead{$f_{\frac{VR}{VR_{max}}}$} & \colhead{$f_{\frac{VR}{VR_{max}}}$ err} & \colhead{S/N}
}
\startdata
53315.33 & -41.9 & 0.562 & 0.843 & 0.007 & 1.501 & 0.67 \\
53323.23 & -35.1 & 0.544 & 0.709 & 0.007 & 1.303 & 0.77 \\
53325.33 & -33.3 & 1.628 & 0.759 & 0.021 & 0.469 & 2.14 \\
53329.32 & -29.8 & 1.489 & 0.940 & 0.019 & 0.633 & 1.58 \\
53344.32 & -16.9 & 3.832 & 1.381 & 0.050 & 0.364 & 2.77 \\
53348.32 & -13.5 & 12.766 & 0.782 & 0.166 & 0.077 & 16.33 \\
53350.33 & -11.7 & 24.416 & 1.230 & 0.318 & 0.069 & 19.85 \\
53356.29 & -6.6 & 57.413 & 0.979 & 0.748 & 0.050 & 58.64 \\
53358.25 & -4.9 & 65.401 & 1.287 & 0.852 & 0.051 & 50.81 \\
53360.25 & -3.2 & 73.437 & 1.385 & 0.957 & 0.051 & 53.01 \\
53377.15 & 11.4 & 44.074 & 0.901 & 0.574 & 0.051 & 48.90 \\
53381.13 & 14.8 & 36.533 & 1.243 & 0.476 & 0.058 & 29.39 \\
53383.13 & 16.5 & 32.523 & 0.863 & 0.424 & 0.054 & 37.68 \\
53387.16 & 20.0 & 24.534 & 0.980 & 0.320 & 0.062 & 25.03 \\
\enddata
\tablecomments{Difference flux lightcurve for SM-2004-LMC-1060. See Table~\ref{tab:2004-LMC-64.tab} for explanation of column headings.}
\label{tab:2004-LMC-1060.tab}
\end{deluxetable}

\begin{deluxetable}{ccccccc}
\tabletypesize{\scriptsize}
\tablecaption{Lightcurve for SM-2004-LMC-1102}
\tablewidth{0pt}
\tablehead{
\colhead{MJD} & \colhead{Rest Phase} & \colhead{Diff flux} & \colhead{flux err} & \colhead{$f_{\frac{VR}{VR_{max}}}$} & \colhead{$f_{\frac{VR}{VR_{max}}}$ err} & \colhead{S/N}
}
\startdata
53315.29 & -40.2 & -0.087 & 0.837 & -0.003 & 9.605 & 0.10 \\
53323.23 & -33.7 & -0.235 & 0.610 & -0.007 & 2.591 & 0.39 \\
53325.33 & -31.9 & -1.052 & 0.728 & -0.033 & 0.693 & 1.45 \\
53329.32 & -28.7 & 0.401 & 0.741 & 0.013 & 1.849 & 0.54 \\
53331.36 & -27.0 & -0.609 & 2.042 & -0.019 & 3.355 & 0.30 \\
53344.32 & -16.4 & 0.643 & 0.836 & 0.020 & 1.301 & 0.77 \\
53348.32 & -13.1 & 6.190 & 0.602 & 0.196 & 0.104 & 10.28 \\
53350.34 & -11.4 & 9.280 & 3.479 & 0.293 & 0.377 & 2.67 \\
53354.28 & -8.2 & 27.495 & 3.617 & 0.869 & 0.137 & 7.60 \\
53356.28 & -6.6 & 25.415 & 1.013 & 0.803 & 0.054 & 25.08 \\
53360.30 & -3.3 & 29.460 & 0.950 & 0.931 & 0.049 & 31.01 \\
53377.17 & 10.5 & 16.716 & 0.786 & 0.528 & 0.060 & 21.27 \\
53381.15 & 13.8 & 12.371 & 0.758 & 0.391 & 0.072 & 16.33 \\
53383.14 & 15.4 & 9.838 & 0.613 & 0.311 & 0.072 & 16.06 \\
53387.17 & 18.7 & 8.589 & 0.611 & 0.271 & 0.080 & 14.05 \\
\enddata
\tablecomments{Difference flux lightcurve for SM-2004-LMC-1102. See Table~\ref{tab:2004-LMC-64.tab} for explanation of column headings.}
\label{tab:2004-LMC-1102.tab}
\end{deluxetable}

\clearpage

\begin{deluxetable}{lcccc}
\tabletypesize{\scriptsize}
\tablecaption{Spectroscopic Observations}
\tablewidth{0pt}
\tablehead{
\colhead{SN ID} & \colhead{Telescope} & \colhead{Instrument} & \colhead{Date} & \colhead{Integration Time (s)} 
}
\startdata
SM-2004-LMC-64 & Magellan II & LDSS-2 & 2004-11-03 & 900 \\
SM-2004-LMC-772 & Magellan I & IMACS-4 & 2004-12-02 & 2400 \\
SM-2004-LMC-797 & Magellan I & IMACS-4 & 2004-12-02 & 2400 \\
SM-2004-LMC-803 & Magellan I & IMACS-4 & 2004-12-02 & 2700 \\
SM-2004-LMC-811 & Magellan I & IMACS-4 & 2004-12-02 & 2700 \\
SM-2004-LMC-917 & Magellan I & IMACS-4 & 2004-12-10 & 1800 \\
SM-2004-LMC-944 & Magellan II & LDSS-2 & 2004-12-18 & 2400 \\
SM-2004-LMC-1002 & Magellan II & LDSS-2 & 2004-12-17 & 2100 \\
SM-2004-LMC-1052 & Magellan II & LDSS-2 & 2004-12-18 & 2400 \\
SM-2004-LMC-1060 & Magellan II & LDSS-2 & 2005-01-11 & 1200 \\
SM-2004-LMC-1102 & Magellan II & LDSS-2 & 2005-01-09 & 3600 \\
\enddata
\label{tab:specinfo}
\end{deluxetable}

\begin{deluxetable}{lccc}

\tabletypesize{\scriptsize} \tablecaption{Redshifts from Galaxy Features} 
\tablewidth{0pt} 
\tablehead{ \colhead{SN ID} & \colhead{galaxy features} & \colhead{galaxy z} & \colhead{SN z}} 
\startdata 
SM-2004-LMC-1002 & CaI H\&K, H$\beta$ & 0.35 & 0.35 \\
SM-2004-LMC-1052 & CaII H\&K, OIII, H$\beta$, OII, H$\gamma$ & 0.348 & 0.34\\
SM-2004-LMC-1060 & CaII H\&K, OIII, H$\beta$, OII, H$\gamma$ & 0.154 & 0.16 \\
\enddata 

\tablecomments{Sources exhibiting strong galactic features.  {\it
SN~ID} indicates the SN whose spectrum shows strong galaxy features.
{\it Galaxy ~features} lists the observed features.  {\it Galaxy~z}
gives the redshift determined from the galaxy lines.  {\it SN~z} gives
the redshift determined through the nearby SN comparison method
described in Section~\ref{section:Spectra}.}

\label{tab:galinfo}
\end{deluxetable}

 \begin{deluxetable}{p{6cm}ccccccccc}
\tabletypesize{\scriptsize}
\tablecaption{Best Fit Parameters}
\tablewidth{0pt}
\rotate
\tablehead{
\colhead{Fit Description} & \colhead{$\chi^{2}/d.o.f.$} & \colhead{$d.o.f.$} & \colhead{$t_{r}$ (days)} & \colhead{$n$ (days)} & \colhead{$m$ (days)} & \colhead{$\tau$ (days)} & \colhead{$\gamma$} & \colhead{$s_{797}$} & \colhead{$s_{803}$}
}
\startdata
Functional Fit (no stretch)\tablenotemark{a} & 1.16 & 38 & -22.2$\pm$0.6 & 9.8$\pm$0.8 & 8.3$\pm$1.4 & 15.2$\pm$0.6 & 0.0029$\pm$0.0005 & \nodata & \nodata\\
Functional Fit (with stretch)\tablenotemark{b} &     0.97 & 36 & -19.2$\pm$1.3 & 13.8$\pm$1.3          & 8.4$\pm$1.5 & 14.0$\pm$0.8 & 0.0030$\pm$0.0006 & 0.92$\pm$0.02 & 1.01$\pm$0.02 \\
\enddata
\tablenotetext{a}{Functional Model of SN~Ia described in Section~\ref{section:templateLC} without stretch parameters to standardize lightcurve width.}
\tablenotetext{b}{Functional Model of SN~Ia with stretch parameters to normalize widths of SNe in composite lightcurve.  Fit normalized to SM-2004-LMC-944.}

\tablecomments{Summary of best fit parameters from fits described in Section~\ref{section:templateLC}.  The fits are performed on the observed $VR$-band $f_{\frac{VR}{VR_{max}}}$ composite lightcurve shifted in time to the SN rest-frame.  The composite lightcurve includes four SNe~Ia with z = 0.135--0.165.  See Section~\ref{section:compositeLC} for description of composite lightcurve construction.} 
\label{tab:parameters}
\end{deluxetable}

\clearpage

\clearpage

\clearpage


\clearpage

\begin{thebibliography}{} 
\bibitem[Ajhar et al.(2001)]{2001ApJ...559..584A} Ajhar, E.~A., Tonry, J.~L., Blakeslee, J.~P., Riess, A.~G., \& Schmidt, B.~P.\ 2001, \apj, 559, 584 
\bibitem[Alard \& Lupton(1998)]{1998ApJ...503..325A} Alard, C., \& Lupton, R.~H.\ 1998, \apj, 503, 325 
\bibitem[Alard(2000)]{2000A&AS..144..363A} Alard, C.\ 2000, \aaps, 144, 363 
\bibitem[Alcock et al.(1999)]{1999ApJ...521..602A} Alcock, C., et al.\ 1999, \apj, 521, 602 
\bibitem[Alcock et al.(2000)]{2000ApJ...542..281A} Alcock, C., et al.\ 2000, \apj, 542, 281 
\bibitem[Aldering et al.(2000)]{2000AJ....119.2110A} Aldering, G., Knop, R., \& Nugent, P.\ 2000, \aj, 119, 2110 
\bibitem[Allington-Smith et al.(1994)]{1994PASP..106..983A} Allington-Smith, J., et al.\ 1994, \pasp, 106, 983
\bibitem[Astier et al. (2006)]{Astier06} Astier, P., et al.\ 2006, \aap\ (in press)
\bibitem[Barris et al. (2005)]{2005AJ....130.2272B} Barris, B.~J., Tonry, J.~L., Novicki, M.~C., \& Wood-Vasey, W.~M.\ 2005, \aj, 130, 2272 
\bibitem[Bigelow et al.(1998)]{1998SPIE.3355..225B} Bigelow, B.~C., Dressler, A.~M., Shectman, S.~A., \& Epps, H.~W.\ 1998, \procspie, 3355, 225 
\bibitem[Bigelow \& Dressler(2003)]{2003SPIE.4841.1727B} Bigelow, B.~C., \& Dressler, A.~M.\ 2003, \procspie, 4841, 1727
\bibitem[Cardelli et al.(1989)]{1989ApJ...345..245C} Cardelli, J.~A., Clayton, G.~C., \& Mathis, J.~S.\ 1989, \apj, 345, 245 
\bibitem[Filippenko(1997)]{1997ARA&A..35..309F} Filippenko, A.~V.\ 1997, \araa, 35, 309 
\bibitem[Gibson et al.(2000)]{2000ApJ...529..723G} Gibson, B.~K., et al.\ 2000, \apj, 529, 723 
\bibitem[G{\"o}ssl \& Riffeser(2002)]{2002A&A...381.1095G} G{\"o}ssl, C.~A., \& Riffeser, A.\ 2002, \aap, 381, 1095 
\bibitem[Goldhaber \& Supernova Cosmology Project Collaboration(1998)]{1998AAS...193.4713G} Goldhaber, G., \& Supernova Cosmology Project Collaboration 1998, Bulletin of the American Astronomical Society, 30, 1325 
\bibitem[Goldhaber et al.(2001)]{2001ApJ...558..359G} Goldhaber, G., et al.\ 2001, \apj, 558, 359 
\bibitem[Hamuy et al.(1996)]{1996AJ....112.2391H} Hamuy, M., Phillips, M.~M., Suntzeff, N.~B., Schommer, R.~A., Maza, J., \& Aviles, R.\ 1996, \aj, 112, 2391 
\bibitem[Hamuy et al.(1996b)]{1996AJ....112.2438H} Hamuy, M., Phillips, M.~M., Suntzeff, N.~B., Schommer, R.~A., Maza, J., Smith, R.~C., Lira, P., \& Aviles, R.\ 1996, \aj, 112, 2438 
\bibitem[Hamuy et al.(1993)]{1993PASP..105..787H} Hamuy, M., Phillips, M.~M., Wells, L.~A., \& Maza, J.\ 1993, \pasp, 105, 787
\bibitem[Harris et al.(1997)]{1997AJ....114.1933H} Harris, J., Zaritsky, D., \& Thompson, I.\ 1997, \aj, 114, 1933 
\bibitem[Kim et al.(1996)]{1996PASP..108..190K} Kim, A., Goobar, A., \& Perlmutter, S.\ 1996, \pasp, 108, 190 
\bibitem[Kowalski et al.(2004)]{2004AAS...205.7115K} Kowalski, M., et al.\ 2004, American Astronomical Society Meeting Abstracts, 205
\bibitem[Lupton \& Monger (1991)]{1991supe.book.....L} Lupton, R., \& Monger, P.\ 1991, Unpublished paper, 1991,  
\bibitem[Matheson et al.(2005)]{2005AJ....129.2352M} Matheson, T., et al.\ 2005, \aj, 129, 2352 
\bibitem[Miknaitis et al.(2005)]{2005AAS...206.4511M} Miknaitis, G., et al.\ 2005, American Astronomical Society Meeting Abstracts, 206
\bibitem[Nugent et al.(2002)]{2002PASP..114..803N} Nugent, P., Kim, A., \& Perlmutter, S.\ 2002, \pasp, 114, 803 
\bibitem[Oestreicher et al.(1995)]{1995A&AS..112..495O} Oestreicher, M.~O., Gochermann, J., \& Schmidt-Kaler, T.\ 1995, \aaps, 112, 495 
\bibitem[Perlmutter et al.(1999)]{1999ApJ...517..565P} Perlmutter, S., et al.\ 1999, \apj, 517, 565 
\bibitem[Phillips(1993)]{1993ApJ...413L.105P} Phillips, M.~M.\ 1993, \apjl, 413, L105 
\bibitem[Poznanski et al.(2002)]{2002PASP..114..833P} Poznanski, D., Gal-Yam, A., Maoz, D., Filippenko, A.~V., Leonard, D.~C., \& Matheson, T.\ 2002, \pasp, 114, 833
\bibitem[Prieto et al.(2006)]{2006astro.ph..3407P} Prieto, J.~L., Rest, A., \& Suntzeff, N.~B.\ 2006, ArXiv Astrophysics e-prints, arXiv:astro-ph/0603407
\bibitem[Rest et al.(2005)]{2005ApJ...634.1103R} Rest, A., et al.\ 2005, \apj, 634, 1103
\bibitem[Riess et al.(1995)]{1995ApJ...438L..17R} Riess, A.~G., Press, W.~H., \& Kirshner, R.~P.\ 1995, \apjl, 438, L17 
\bibitem[Riess et al.(1996)]{1996ApJ...473...88R} Riess, A.~G., Press, W.~H., \& Kirshner, R.~P.\ 1996, \apj, 473, 88 
\bibitem[Riess et al.(1996b)]{1996ApJ...473..588R} Riess, A.~G., Press, W.~H., \& Kirshner, R.~P.\ 1996, \apj, 473, 588 
\bibitem[Riess et al.(1998)]{1998AJ....116.1009R} Riess, A.~G., et al.\ 1998, \aj, 116, 1009
\bibitem[Riess et al.(1999)]{1999AJ....118.2675R} Riess, A.~G., et al.\ 1999, \aj, 118, 2675 
\bibitem[Schlegel et al.(1998)]{1998ApJ...500..525S} Schlegel, D.~J., Finkbeiner, D.~P., \& Davis, M.\ 1998, \apj, 500, 525 
\bibitem[Schmidt et al.(1998)]{1998ApJ...507...46S} Schmidt, B.~P., et al.\ 1998, \apj, 507, 46 
\bibitem[Stubbs et al.(2002)]{2002AAS...201.7807S} Stubbs, C.~W., et al.\ 2002, Bulletin of the American Astronomical Society, 34, 1232 
\bibitem[Suntzeff et al.(1999)]{1999AJ....117.1175S} Suntzeff, N.~B., et al.\ 1999, \aj, 117, 1175 
\bibitem[van Dokkum(2001)]{2001PASP..113.1420V} van Dokkum, P.~G.\ 2001, \pasp, 113, 1420
\bibitem[Zaritsky et al.(2004)]{2004AJ....128.1606Z} Zaritsky, D., Harris, J., Thompson, I.~B., \& Grebel, E.~K.\ 2004, \aj, 128, 1606 
\end{thebibliography}
\end{document}